\documentclass[preprintnumbers,noshowpacs,prl,10pt,twocolumn,floatfix]{revtex4}%
\pdfoutput=1
\usepackage{amssymb}
\usepackage{amsfonts}
\usepackage{amsmath}
\usepackage{graphicx}
\usepackage{dcolumn}
\usepackage{bm}%
\begin{document}
\title{Impulse-induced optimum control of chaos in dissipative driven systems}
\author{Pedro J. Mart\'{\i}nez$^{1\ast}$, Stefano Euzzor$^{2}$, Jason A. C. Gallas$^{3}$,
Riccardo Meucci$^{2}$ and Ricardo Chac\'{o}n$^{4}$}
\affiliation{$^{1}$Departamento de F\'{\i}sica Aplicada, E.I.N.A., Universidad de Zaragoza,
E-50018 Zaragoza, Spain, and Instituto de Ciencia de Materiales de Arag\'{o}n,
CSIC-Universidad de Zaragoza, E-50009 Zaragoza, Spain}
\affiliation{$^{2}$Istituto Nazionale di Ottica, Consiglio Nazionale delle Ricerche, Largo
E. Fermi 6, Firenze, Italy}
\affiliation{$^{3}$Departamento de F\'{\i}sica, Universidade Federal da Para\'{\i}ba,
58051-970 Joao Pessoa, Brazil}
\affiliation{$^{4}$Departamento de F\'{\i}sica Aplicada, E.I.I., Universidad de
Extremadura, Apartado Postal 382, E-06006 Badajoz, Spain, and Instituto de
Computaci\'{o}n Cient\'{\i}fica Avanzada, Universidad de Extremadura, E-06006
Badajoz, Spain}
\date{\today}

\begin{abstract}
Taming chaos arising from dissipative non-autonomous nonlinear systems by
applying additional harmonic excitations is a reliable and widely used
procedure nowadays. But the suppressory effectiveness of generic non-harmonic
periodic excitations continues to be a significant challenge both to our
theoretical understanding and in practical applications. Here we show how the
effectiveness of generic suppressory excitations is optimally enhanced when
the impulse transmitted by them (time integral over two consecutive zeros) is
judiciously controlled in a not obvious way. This is demonstrated
experimentally by means of an analog version of a universal model, and
confirmed numerically by simulations of such a damped driven system including
the presence of noise. Our theoretical analysis shows that the controlling
effect of varying the impulse is due to a correlative variation of the energy
transmitted by the suppressory excitation.

\end{abstract}
\maketitle

Obtaining full control of the chaotic dynamics of generic dissipative
non-linear systems represents a fundamental interdisciplinary scientific and
technological challenge. Among the different control procedures which have
been proposed [1-3], the application of judiciously chosen periodic
excitations [4-20] constitutes a reliable procedure in the context of
dissipative non-autonomous systems. Hitherto, experimental control of chaos by
periodic excitations has been demonstrated in many diverse systems, including
laser systems [8,10,13,16], neurological systems [11], ferromagnetic systems
[5], chemical reactions [17], and electronic systems [7,20]. It has been shown
that the effectiveness of this non-feedback control procedure in
non-autonomous systems depends critically upon the resonance condition and the
initial phase difference between the primary (or chaos-inducing) periodic
excitation and the secondary (or suppressory) periodic excitation, which has
given rise to its denomination as phase control [19,20]. In such previous
works, however, the flexibility of the control scenario against diversity in
the suppressory excitations (SEs) was not studied since harmonic excitations
have been overwhelmingly considered for the compelling reason of their
simplicity. Clearly, the assumption of harmonic excitations means that the
driving systems$-$whatever they might be$-$are effectively taken as linear.
This mathematically convenient choice imposes a drastic and unnecessary
restriction in the control scenario which is untenable for most natural and
artificial systems due to their irreducible nonlinear nature [21]. Thus, to
fully explore and exploit the physics of the control scenario, it seems
appropriate to consider SEs exhibiting general features of periodic
excitations which are the output of nonlinear systems, therefore being
appropriately represented by Fourier series$-$not just by a single harmonic
term. It has been shown, in particular, that the suppressory effectiveness of
periodic excitations seems to be highly sensitive to their wave forms [2].
Since there are infinitely many different waveforms, an important question,
both scientifically and technologically, is how can one explain in physical
terms$-$providing in turn a quantitative characterization$-$the effect of the
SE's waveform on the control scenario.

Here, we experimentally demonstrate that a relevant quantity properly
characterizing the effectiveness of generic SEs $f(t)$ having equidistant
zeros in the control scenario is the \textit{impulse} transmitted by the
excitation over a half-period (hereafter referred to simply as the
excitation's impulse,
\begin{equation}
I\equiv\int_{0}^{T/2}f(t)dt, \tag{1}%
\end{equation}
with $T$ being the period)$-$ a quantity integrating the conjoint effects of
the excitation's amplitude, period, and waveform. The relevance of the
excitation's impulse has been observed previously in such different contexts
as adiabatically ac-driven periodic Hamiltonian systems [22], chaotic dynamics
of lasers [23], and discrete soliton ratchets [24], to cite just a few
instances. For the sake of clarity, we consider an analog implementation of a
simple universal model to discuss the impulse-induced chaos-control scenario:
A damped-driven two-well Duffing oscillator described by the equation:
\begin{equation}
\overset{..}{x}=x-\beta\left[  1+\eta f(t)\right]  x^{3}-\delta\overset{.}%
{x}+\gamma\cos\left(  \omega t\right)  , \tag{2}%
\end{equation}
where all the variables and parameters are dimensionless $\left(  \beta
,\eta,\delta,\gamma>0\right)  $, while $f(t)$ is an unit-amplitude
$T$-periodic excitation chosen to satisfy three remarkable properties. First,
its waveform (and hence its impulse) is changed by solely varying a
\textit{single} parameter, the shape parameter $m$, between 0 and 1. Second,
when $m=0$, then $f\left(  t\right)  _{m=0}=\sin\left(  2\pi t/T+\varphi
\right)  $, with $\varphi$ being the initial phase difference between the two
excitations involved for all values of the shape parameter, i.e., one recovers
the standard case [20] of an harmonic excitation, while for the limiting value
$m=1$ the excitation and its impulse vanish. And third, as a function of $m$,
the SE's impulse presents a single maximum at a certain value $m=m_{\max
}^{impulse}$ (see Fig.~1 and [25] for the definition and additional properties
of $f(t)$). Here, $\gamma\cos\left(  \omega t\right)  $ and $-\beta\eta
x^{3}f\left(  t\right)  $ are to be regarded for convenience as the primary
and suppressory excitations, respectively. 
\begin{figure}
\includegraphics[width=0.45\textwidth]{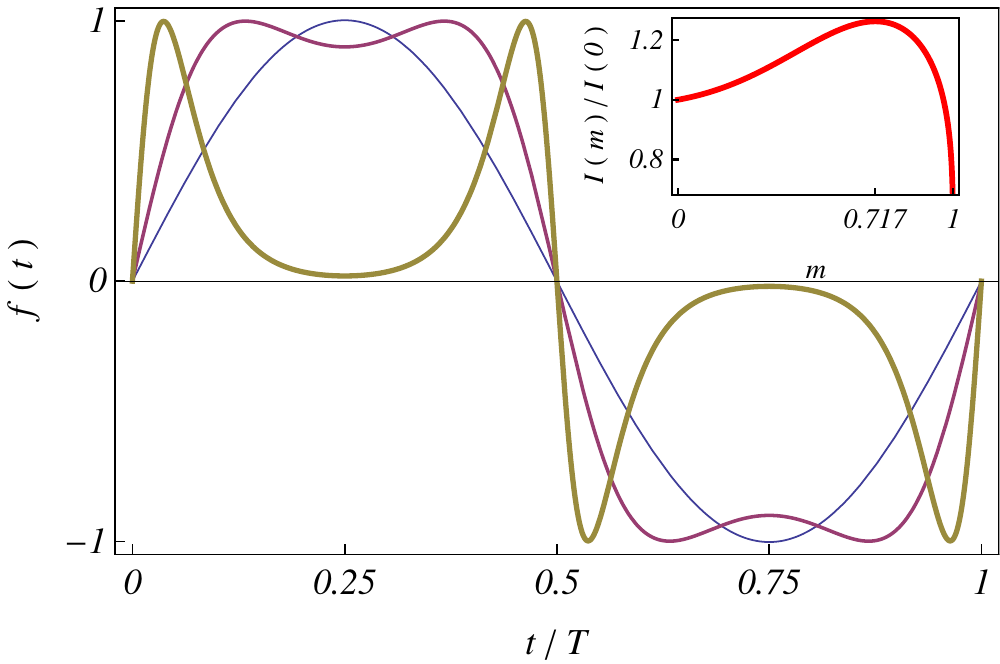}
\caption{Suppressory $T$-periodic excitation $f(t)$ versus
$t/T$ for three
values of the shape parameter: $m=0$ (sinusoidal pulse, thin line),
$m=0.717\simeq m_{\max}^{impulse}$ (nearly square-wave pulse, medium
line),
and $m=0.9999$ (double-humped pulse, thick line). The inset shows the
corresponding normalized impulse $I(m)/I(m=0)$ versus $m$.
}
\label{fig1}
\end{figure}

Also, we assume that, in the
absence of any SE $\left(  \eta=0\right)  $, the Duffing oscillator (2)
presents steady chaotic behavior which ultimately comes from a homoclinic
bifurcation [26], while we will focus here on the effective case of the main
resonance $\left(  T=2\pi/\omega\right)  $ between the two involved
excitations in the presence of SEs $\left(  \eta>0\right)  $. As shown below,
the simple and natural choice for $f(t)$ allows us to characterize
experimentally the genuine effect on the chaos-control scenario of the impulse
transmitted by \textit{generic} SEs, as well as to explain theoretically that
the controlling effect of varying the impulse is due to a correlative
variation of the energy transmitted by the SE, allowing us to obtain useful
analytical estimates of the chaotic threshold in the $\varphi-\eta$ control
plane from Melnikov [26] and energy-based analyses, as is detailed in the
Supplemental Material [25].

We investigated the impulse-induced chaos-control scenario in the laboratory
by implementing an analog version of the Duffing oscillator (2) (see [25] for
additional details). Our experimental results systematically indicate that
complete regularization (i.e., periodic responses of any periodicity order)
mainly appears inside two maximal islands in the $\varphi-\eta$ control plane
which are roughly symmetric with respect to the two optimal suppressory values
$\varphi_{opt}\equiv\left\{  \pi/2,3\pi/2\right\}  $, respectively, for all
values of the shape parameter (see Fig. 2).

\begin{figure}
\includegraphics[width=0.5\textwidth]{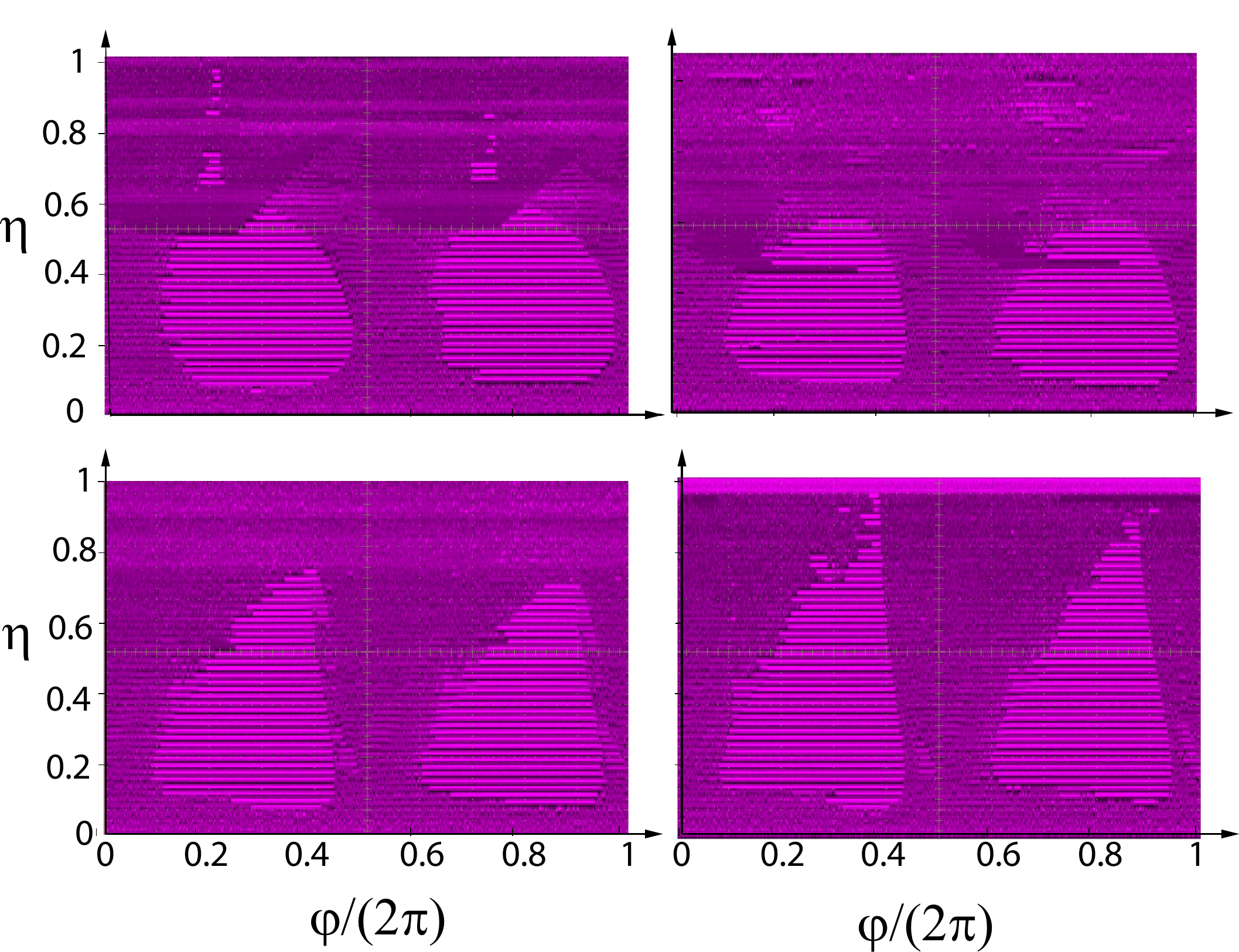}
\caption{ Experimentally obtained regions in the $\varphi-\eta$ control plane
with $\varphi\in\left[  0,2\pi\right]  $ and $\eta\in\left[  0,1\right]  $
corresponding to chaos (non-uniform magenta regions), low-energy periodic
orbits around some of the two fixed points $\left(  x=\pm\beta^{-1/2}%
,\overset{.}{x}=0\right)  $ of the unperturbed Duffing oscillator (uniform
light magenta regions), and higher-energy periodic orbits encircling both
fixed points (uniform dark magenta regions) for four values of the shape
parameter: (a) $m=0$, (b) $m=0.717\simeq m_{\max}^{impulse}$, (c) $m=0.9$, and
(d) $m=0.95$. Fixed parameters: $\delta=0.25,\gamma=0.29,\beta=1,\omega=1$.
}
\label{fig2}
\end{figure}

The analysis of the experimental
data gives rise to the following genuine features of the present chaos-control scenario.
While both the size and the form of the boundaries of the maximal
regularization islands vary as the SE's impulse changes by solely varying $m$,
they remain roughly centered around the optimal values $\varphi_{opt}%
\equiv\left\{  \pi/2,3\pi/2\right\}  $ (note that the entire diagrams of Fig.
2 are periodic along the $\varphi$-axis, with fundamental period equal to
$\pi$), confirming thus the theoretical predictions from Melnikov and
energy-based analyses [25].

\begin{figure}
\includegraphics[width=0.5\textwidth]{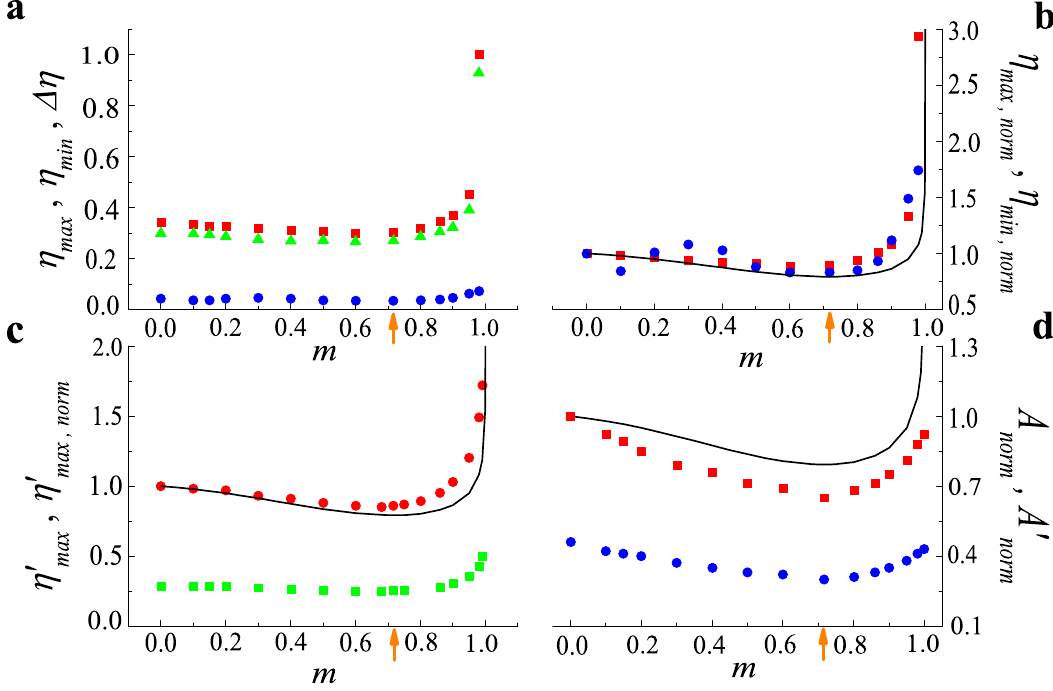}
\caption{ Experimental values of threshold amplitudes and regularization area in
the control parameter plane versus shape parameter: (a) Lower threshold
amplitude $\eta_{\min}$ (circles), upper threshold amplitude $\eta_{\max}$
(squares), and difference $\Delta\eta\equiv\eta_{\max}-\eta_{\min}$
(triangles) versus shape parameter $m$. (b) Normalized lower threshold
amplitude $\eta_{\min,norm}=\eta_{\min,norm}\left(  m\right)  \equiv\eta
_{\min}(m)/\eta_{\min}(m=0)$ (circles), normalized upper threshold amplitude
$\eta_{\max,norm}=\eta_{\max,norm}(m)\equiv\eta_{\max}(m)/\eta_{\max}(m=0)$
(squares), and inverse of the normalized impulse $\left[  I(m)/I(m=0)\right]
^{-1}$ (solid line; cf. Eq. (S3) in Supplemental Material [25]). (c) Threshold
amplitude $\eta_{\max}^{\prime}$ leading the Duffing oscillator to
small-amplitude periodic oscillations around one of the fixed points $\left(
x=\pm\beta^{-1/2},\overset{.}{x}=0\right)  $ of the unperturbed Duffing
oscillator (squares), its normalized version $\eta_{\max,norm}^{\prime}%
=\eta_{\max,norm}^{\prime}(m)\equiv\eta_{\max}^{\prime}(m)/\eta_{\max}%
^{\prime}(m=0)$ (circles), and analytical estimate of the latter [solid line;
cf. Eq. (3)]. (d) Normalized areas of regularized regions in the $\varphi
-\eta$ control plane, $A_{norm}=A_{norm}(m)\equiv A(m)/A\left(  m=0\right)  $
(squares), $A_{norm}^{\prime}=A_{norm}^{\prime}\left(  m\right)  \equiv
A(m)/A_{total}$ (circles), in which $A(m)$ and $A_{total}$ are the
regularization area and the total area, respectively. The solid line denotes
the inverse of the normalized impulse $\left[  I(m)/I(m=0)\right]  ^{-1}$,
whereas the orange arrows indicate the value $m=m_{\max}^{impulse}\simeq
0.717$, i.e., the $m$ value at which the SE's impulse is maximum.  Fixed
parameters: $\delta=0.25,\gamma=0.29,\beta=1,\omega=1$.
}
\label{fig3}
\end{figure}

The lower, $\eta_{\min}$, and upper, $\eta_{\max}$, threshold values of the
SE's amplitude measured at the optimal suppressory values $\varphi
=\varphi_{opt}\equiv\left\{  \pi/2,3\pi/2\right\}  $ as well as the difference
$\Delta\eta\equiv\eta_{\max}-\eta_{\min}$ present, as functions of the shape
parameter, a behavior quite similar to that of the inverse of the SE' impulse
[see Fig. 3(a)]. This can be seen more clearly in Fig. 3(b) in which it is
shown the normalized amplitude thresholds $\eta_{\max}(m)/\eta_{\max}(m=0)$,
$\eta_{\min}(m)/\eta_{\min}(m=0)$ together with the inverse of the normalized
impulse $\left[  I(m)/I(m=0)\right]  ^{-1}$ for the sake of comparison (see
Supplemental Material [25]). In particular, we can see that the respective
minima occur at values of the shape parameter which are very close in the
sense that the difference between the corresponding values of the SE's impulse
is hardly noticeable. 

Although we have not gotten a definitive explanation of
the apparently anomalous behavior of $\eta_{\min}$ over a certain range of
\textit{small} values of $m$, it seems to be originated in the fractal
character of the boundary for chaos in parameter space [27] together with the
fact that over such a range of $m$ values the changes of the SE's impulse are
hardly noticeable [25]. The experimental results shown in Fig. 3(a) indicate
that ever lower amplitudes $\eta_{\min}$ can suppress chaos as the impulse
transmitted by the SE approaches its maximum value, whereas the corresponding
suppressory ranges $\Delta\eta$ also decrease in the same way as $\eta_{\min}$
owing to the impulse-induced \textit{enhancement} of the chaos-inducing
effectiveness of the SE. This dependence of $\eta_{\min},\eta_{\max}%
,\Delta\eta$ on the SE's impulse, which is theoretically anticipated from
Melnikov analysis [25], represents an essential feature of the present
chaos-control scenario which is expected to be independent of the particular
choice for the SE.

The lower values of the SE's amplitude which suppress chaos by leading the
Duffing oscillator to small-amplitude periodic oscillations around one of the
fixed points $\left(  x=\pm\beta^{-1/2},\overset{.}{x}=0\right)  $ of the
unperturbed Duffing oscillator $\left(  \delta=\gamma=\eta=0\right)  $,
$\eta_{\max}^{\prime}$, present, as a function of the shape parameter, a
behavior quite similar to that of the inverse of the SE's impulse [see Fig.
3(c)]. Remarkably, we can see in Fig. 3(c) that the theoretical estimate of
its normalized version,
\begin{equation}
\frac{\eta_{\max}^{\prime}(m)}{\eta_{\max}^{\prime}(m=0)}=\left[  \frac
{I(m)}{I(m=0)}\right]  ^{-1}, \tag{3}%
\end{equation}
fits quite well the corresponding experimental values. Since the energy-based
analysis giving rise to Eq. (3) is \textit{general} in the sense that it can
be applied to damped-driven systems of type (1) with generic (analytical)
potentials $U(x)$ (see Supplemental Material [25]), one may expect that the
dependence of $\eta_{\max}^{\prime}$ on the SE's impulse represents an
additional universal feature of the present chaos-control scenario.

The total area of regularized regions (i.e., those associated with periodic
responses of any periodicity order), $A$, in the $\varphi-\eta$ control plane
presents, as a function of the shape parameter, a behavior which exhibits
relevant features that are common to those of the inverse of the SE's impulse.
Specifically, Fig. 3(d) shows that its normalized versions $A_{norm}\equiv
A(m)/A\left(  m=0\right)  $ and $A_{norm}^{\prime}\equiv A(m)/A_{total}$
present a single minimum just at $m=m_{\max}^{impulse}\simeq0.717$, i.e., the
$m$ value at which the SE's impulse is maximum (see Fig.~1). It is worth
noting that the same behavior is theoretically anticipated for the area of the
aforementioned maximal islands from the application of the Melnikov analysis
to the crudest approximation of the SE $f(t)$, i.e., when solely the main
harmonic of its Fourier expansion is retained (see Supplemental Material [25]
for an analytical estimate of the maximal islands' area). This
\textit{inverse} dependence of the regularization areas in the $\varphi-\eta$
control plane on the SE's impulse represents an additional essential feature
of the present chaos-control scenario which is expected to be especially
useful in technological applications owing to it provides an universal
criterion to guide the design of optimal SEs.

\begin{figure}
\includegraphics[width=0.5\textwidth]{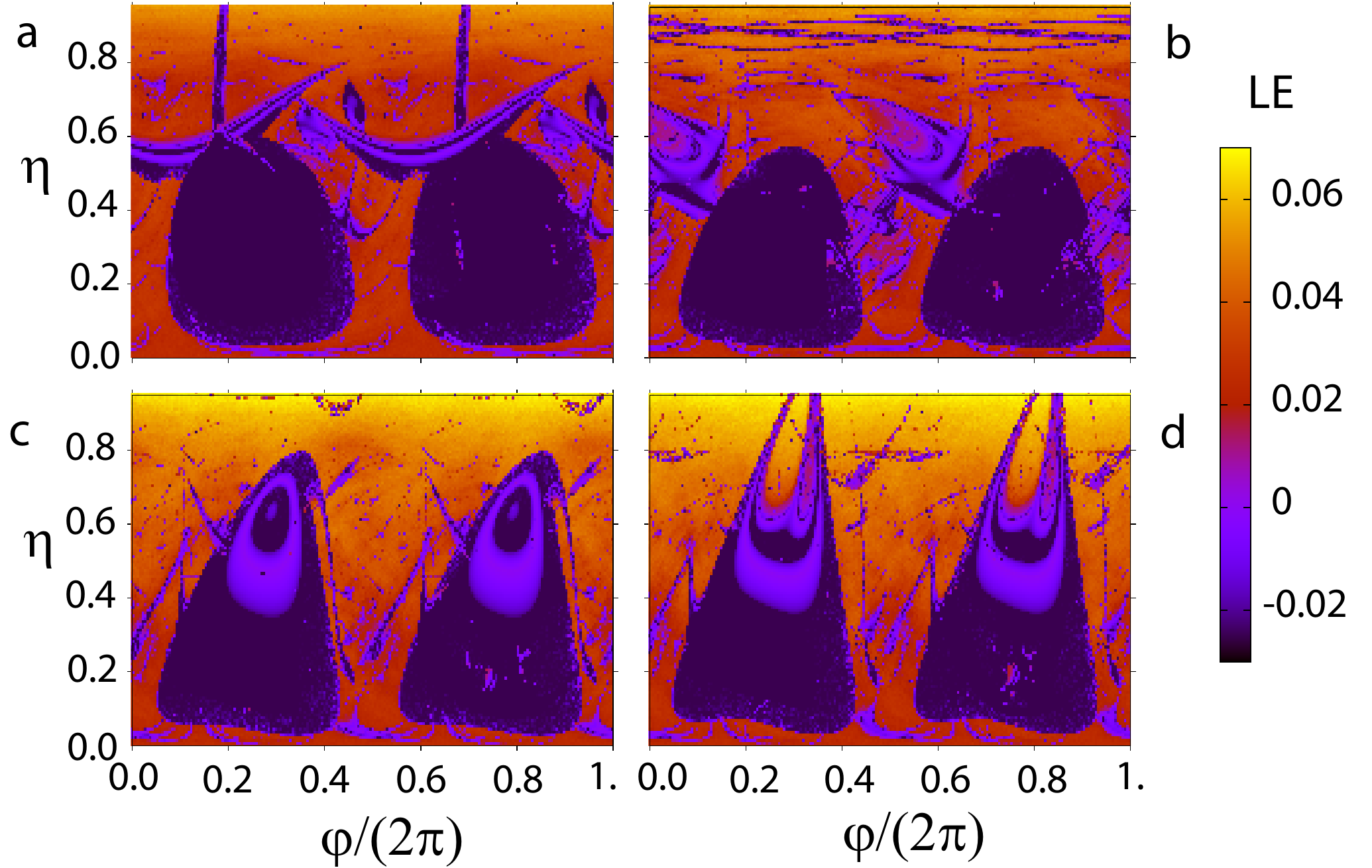}
\caption{ Numerically calculated maximal Lyapunov exponent in the $\varphi-\eta$
control plane for four values of the shape parameter: (a) $m=0$, (b)
$m=0.717\simeq m_{\max}^{impulse}$ (i.e., the $m$ value at which the SE's
impulse is maximum), (c) $m=0.9$, and (d) $m=0.95$. Fixed parameters:
$\delta=0.25,\gamma=0.29,\beta=1,\omega=1$.
}
\label{fig4}
\end{figure}

Extensive computer simulations of Eq. (1) yielded numerical results from which
we constructed three complementary types of diagrams providing useful
information on both regularization regions in the $\varphi-\eta$ control plane
and the nature of the regularized (periodic) responses: maximal Lyapunov
exponent, period-distribution, and isospike diagrams (see Supplemental
Material [25]). The conclusions arising from the analysis of these diagrams
systematically agree with all the aforementioned experimental features of the
present chaos-control scenario, as can be appreciated by comparing the maximal
Lyapunov exponent diagrams shown in Fig. 4 with the respective experimental
diagrams shown in Fig. 3. 

\begin{figure}
\includegraphics[width=0.45\textwidth]{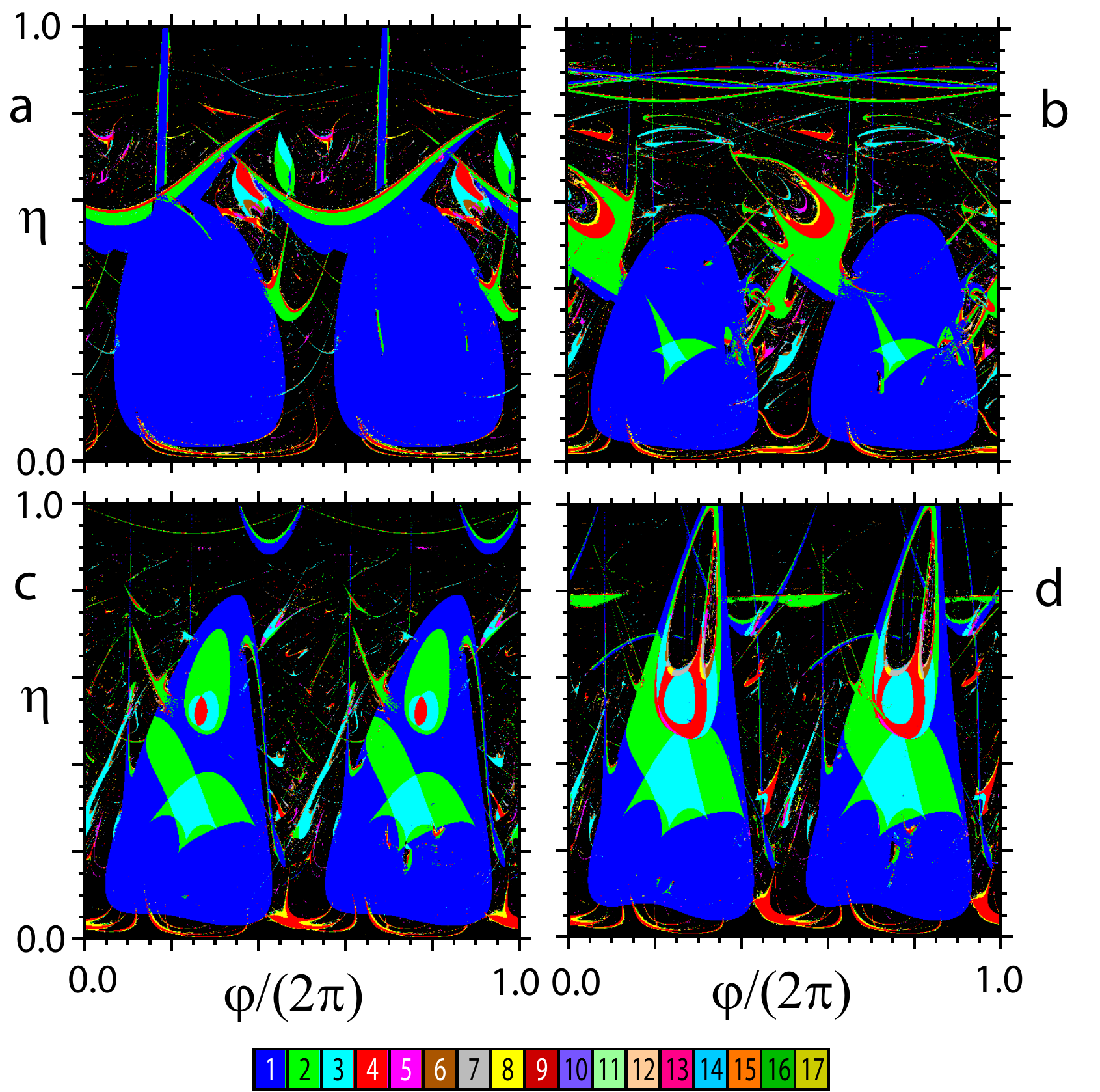}
\caption{ Numerically calculated regularization regions according to the
waveform complexity (number of spikes or local maxima per period) of their
solutions and chaotic regions (black) in the $\varphi-\eta$ control plane for
four values of the shape parameter: (a) $m=0$, (b) $m=0.717\simeq m_{\max
}^{impulse}$ (i.e., the $m$ value at which the SE's impulse is maximum), (c)
$m=0.9$, and (d) $m=0.95$. Fixed parameters: $\delta=0.25,\gamma
=0.29,\beta=1,\omega=1$.
}
\label{fig5}
\end{figure}

Regarding the nature of the regularized responses,
the period-distribution and isospike diagrams inform us of the existence of a
wide spectrum of periodic responses in different regions of the $\varphi-\eta$
control plane, the period-1 solutions being the prevailing responses over the
two maximal regularization islands irrespective of the values of the SE's
impulse (see Figs. 5 and 6). 

\begin{figure}
\includegraphics[width=0.45\textwidth]{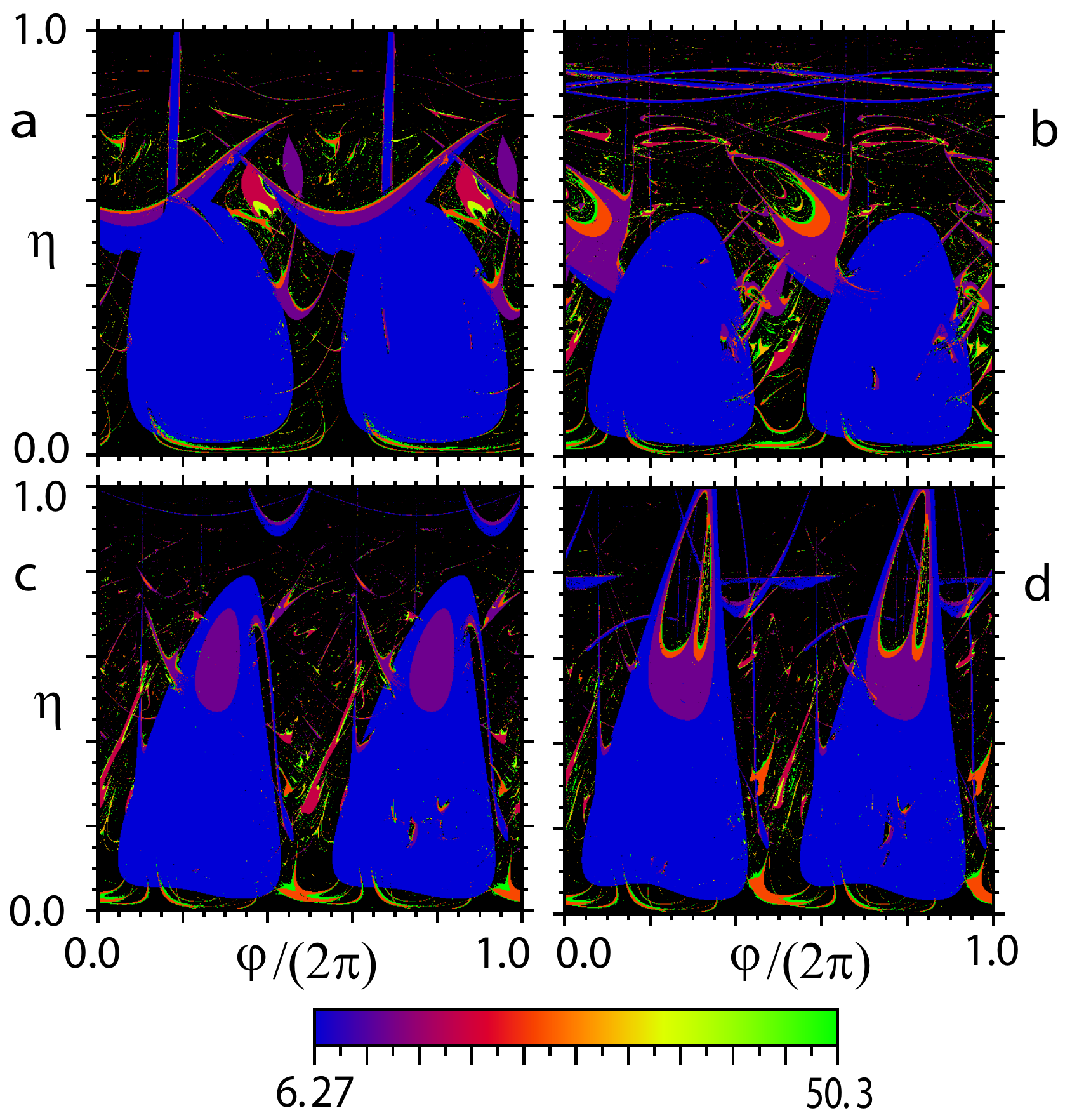}
\caption{ Numerically calculated regularization regions according to the period
of their periodic solutions and chaotic regions (black) in the $\varphi-\eta$
control plane for four values of the shape parameter: (a) $m=0$, (b)
$m=0.717\simeq m_{\max}^{impulse}$ (i.e., the $m$ value at which the SE's
impulse is maximum), (c) $m=0.9$, and (d) $m=0.95$. Fixed parameters:
$\delta=0.25,\gamma=0.29,\beta=1,\omega=1$.
}
\label{fig6}
\end{figure}

Importantly, our numerical results show that the
present chaos-control scenario is robust against the presence of
moderate-intensity Gaussian noise, with the two maximal regularization islands
being the robustest regularization regions, which represents an invaluable
feature due to the unavoidable presence of thermal noise in many physical
contexts, including for instance many nanoscale devices. Specific examples are
shown in [25].

During the last three decades or so [1-3], and on the basis of an overwhelming
use of harmonic SEs, the effectiveness of this particular type of SE has been
systematically explored in a vast diversity of physical contexts by
independently varying its amplitude and frequency as control parameters.
However, by taking into account the irreducible nonlinear nature of real-world
periodic excitations, the present results demonstrate that the SE's impulse is
the relevant quantity providing a complete characterization of the suppressory
effectiveness of generic SEs by means of an exquisite control of the injection
of energy into a chaotic damped-driven system. Future work may extend the
present impulse-induced chaos-control scenario to the control of diverse
quantum phenomena associated with the so-called quantum chaos, such as
dynamical localization [28] and quantum entanglement in systems in contact
with environment [29].


\widetext
\clearpage
\begin{center}
\textbf{\large Supplemental Material: Impulse-induced optimum control of chaos in dissipative driven systems}
\end{center}

\setcounter{equation}{0}
\setcounter{figure}{0}
\setcounter{page}{1}
\makeatletter
\renewcommand{\bibnumfmt}[1]{[S#1]}
\renewcommand{\citenumfont}[1]{S#1}
\renewcommand{\thefigure}{S\arabic{figure}}

{\LARGE \bigskip}

\section{Theoretical methods}

\subsection{Fourier expansion of the suppressory excitation (SE)}

In our study we consider the elliptic SE
\begin{equation}
f(t)\equiv N\operatorname*{sn}\left(  4Kt/T+\Phi\right)  \operatorname*{dn}%
\left(  4Kt/T+\Phi\right)  , \tag{S1}%
\end{equation}
in which $\operatorname*{sn}\left(  \cdot\right)  \equiv\operatorname*{sn}%
\left(  \cdot;m\right)  $ and $\operatorname*{dn}\left(  \cdot\right)
\equiv\operatorname*{dn}\left(  \cdot;m\right)  $ are Jacobian elliptic
functions of parameter $m$ ($K\equiv K(m)$ is the complete elliptic integral
of the first kind) \cite{S_1}, $\Phi=\Phi\left(  m,\varphi\right)  \equiv
2K(m)\varphi/\pi$, $\varphi\in\left[  0,2\pi\right]  $, $T\equiv2\pi/\omega$,
and
\begin{equation}
N=N(m)\equiv\left[  a+b\left(  1+\exp\left\{  \frac{m-c}{d}\right\}  \right)
^{-1}\right]  ^{-1}, \tag{S2}%
\end{equation}
is a normalization function ($a\equiv0.43932,b\equiv0.69796,c\equiv
0.3727,d\equiv0.26883$) which is introduced for the elliptic excitation to
have the same amplitude, 1, and period $T$, for any waveform (i.e., $\forall
m\in\left[  0,1\right]  $). When $m=0$, then $f\left(  t\right)  _{m=0}%
=\sin\left(  2\pi t/T+\varphi\right)  $, i.e., one recovers the standard case
of an harmonic SE, while for the limiting value $m=1$ the excitation vanishes.
The effect of renormalization of the elliptic arguments is clear: with $T$
constant, solely the excitation's impulse is varied by increasing the shape
parameter $m$ from $0$ to $1$. Note that, as a function of $m$, the SE's
impulse per unit of amplitude and unit of period
\begin{equation}
I=I(m)\equiv\frac{N\left(  m\right)  }{2K\left(  m\right)  } \tag{S3}%
\end{equation}
presents a single maximum at $m=m_{\max}^{impulse}\simeq0.717$ (see Fig. 1 of
the main text).

The Fourier expansion of the elliptic SE (Eq. S1) reads%
\begin{align}
f(t)  &  =\sum_{n=0}^{\infty}a_{n}(m)\sin\left[  \left(  2n+1\right)  \left(
\frac{2\pi t}{T}+\varphi\right)  \right]  ,\tag{S4}\\
a_{n}(m)  &  \equiv\frac{\pi^{2}N(m)(n+\frac{1}{2})}{\sqrt{m}K^{2}%
(m)}\operatorname{sech}\left[  \frac{(n+\frac{1}{2})\pi K(1-m)}{K(m)}\right]
, \tag{S5}%
\end{align}
in which its Fourier coefficients satisfy the properties: i) $\lim
_{m\rightarrow1}a_{n}(m)=0$, ii) $a_{n}(m)$ exhibits a single maximum at
$m=m_{\max}\left(  n\right)  $ such that $m_{\max}\left(  n+1\right)
>m_{\max}\left(  n\right)  $, $n=0,1,...$, iii) the normalized functions
$a_{0}(m)/a_{0}(m=0)$ and $I(m,T)/I(m=0,T)\equiv\pi N(m)/(2K(m))$ present, as
functions of $m$, similar behaviours while their maxima verify that $m_{\max
}\left(  n=0\right)  \simeq0.65$ is very close to $m_{\max}^{impulse}%
\simeq0.717$ (see Fig. S1), and iv) the Fourier expansion (Eq. S4) is rapidly
convergent over a wide range of values of the shape parameter. The following
remarks may now be in order.

\begin{figure}
\includegraphics[width=0.55\textwidth]{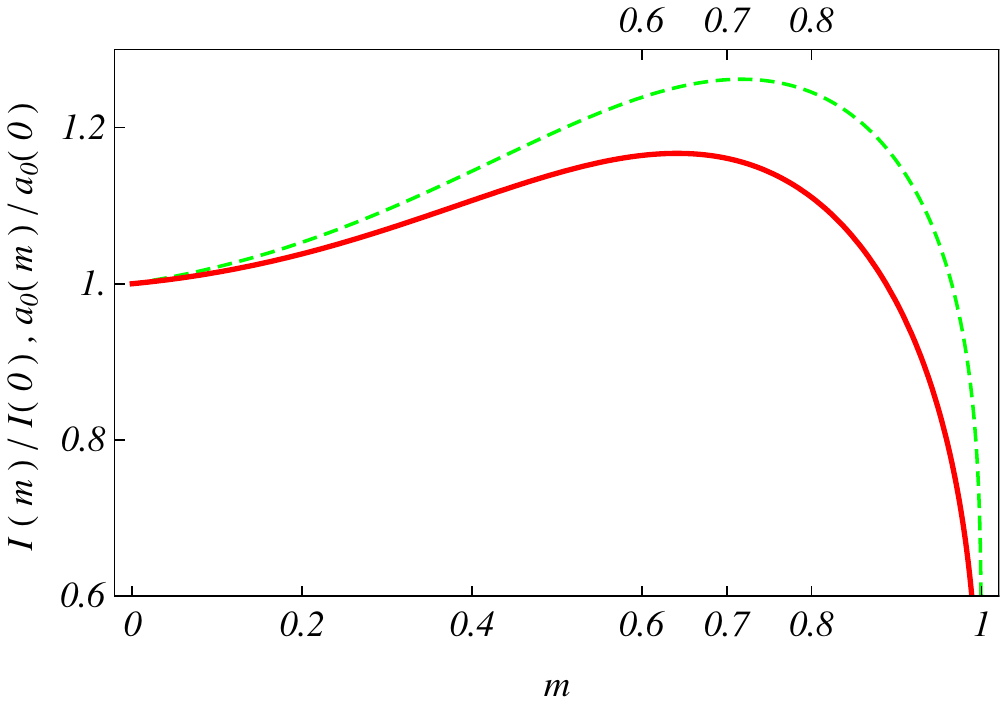}
\caption{\textbf{Comparison between the SE's impulse and its first
Fourier
coefficient as functions of the shape parameter.} Normalized first
Fourier
coefficient $a_{0}(m)/a_{0}(m=0)$ (Eq. S5, solid line) and SE's impulse
$I(m)/I(m=0)\equiv\pi N(m)/(2K(m))$ (Eq. S3, dashed line) versus shape
parameter $m$. We can see that the respective single maxima occur at
very
close values of the shape parameter: $m_{\max}\left(  n=0\right)
\simeq0.642$
and $m_{\max}^{impulse}\simeq0.717$, respectively.
}
\label{figS1}
\end{figure}

First, regarding analytical estimates, the property (iii) is relevant in the
sense that it allows us to obtain an useful effective estimate of the chaotic
threshold in the $\varphi-\eta$ control plane from Melnikov analysis (MA)
\cite{S_2,S_3} by solely retaining the first harmonic of the Fourier expansion (Eq.
S4):
\begin{equation}
f(t)\approx a_{0}(m)\sin\left(  \omega t+\varphi\right)  . \tag{S6}%
\end{equation}

Second, regarding experiments, the property (iv) is relevant in the sense that
it allows us to effectively approximate the elliptic SE by solely retaining
the first two harmonics of its Fourier expansion over the range of values of
the shape parameter of our interest ($0\leqslant m\lesssim0.95$; see Fig. S2):%
\begin{equation}
f(t;T,m,\varphi)\approx a_{0}(m)\sin\left(  \omega t+\varphi\right)
+a_{1}(m)\sin\left(  3\omega t+3\varphi\right)  . \tag{S7}%
\end{equation}
\begin{figure}
\includegraphics[width=0.75\textwidth]{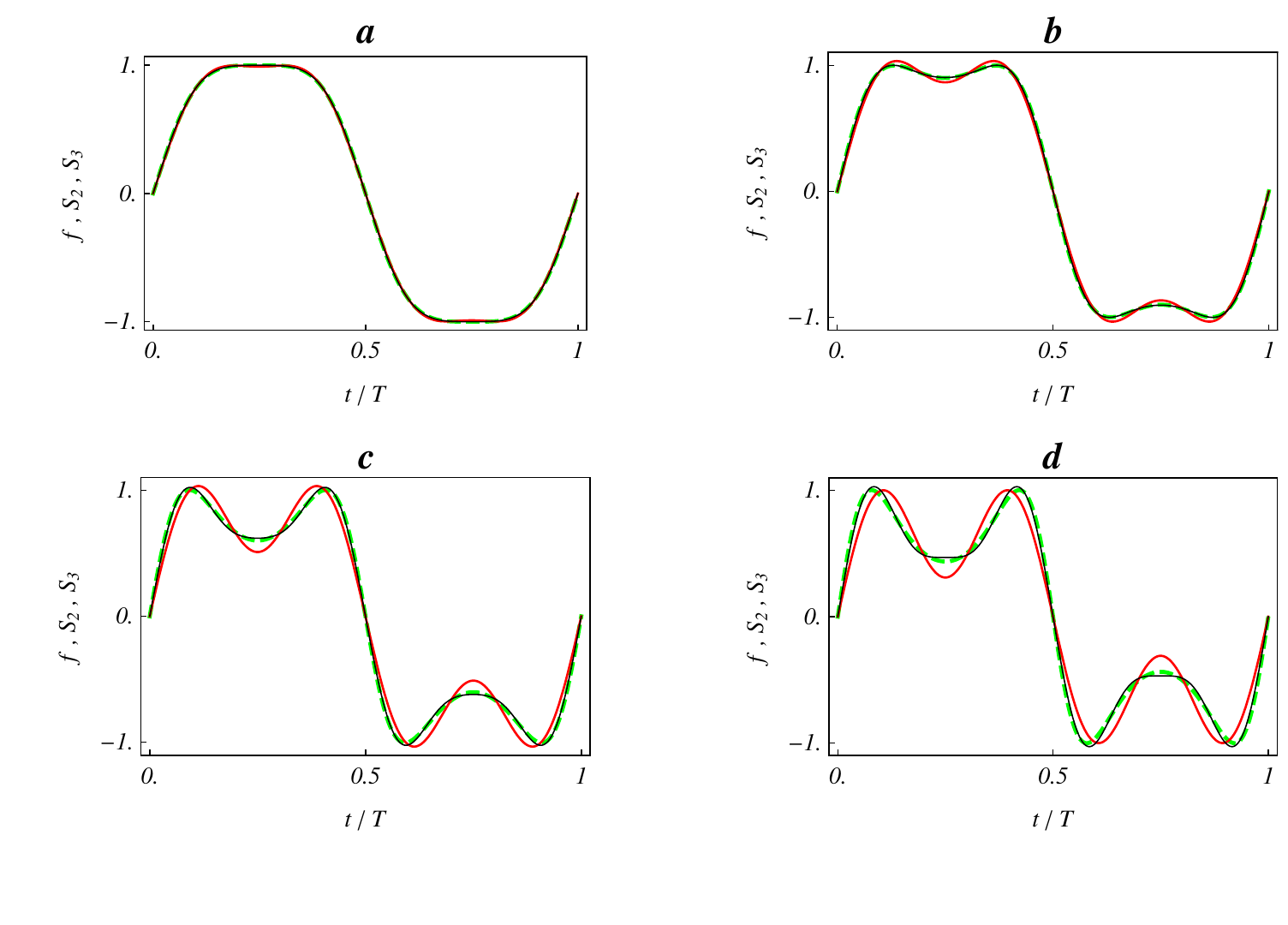}
\caption{\textbf{Comparison between the elliptic SE and its
two- and
three-harmonics approximations over a period for four values of the
shape
parameter.} Plots of the elliptic SE (Eq. S1, dashed line), its
two-harmonics
approximation $S_{2}(t)\equiv a_{0}(m)\sin\left(  \omega
t+\varphi\right)
+a_{1}(m)\sin\left(  3\omega t+3\varphi\right)  $ (cf. Eqs. S4 and S5,
solid
line), and its three-harmonics approximation $S_{3}(t)\equiv a_{0}%
(m)\sin\left(  \omega t+\varphi\right)  +a_{1}(m)\sin\left(  3\omega
t+3\varphi\right)  +a_{2}(m)\sin\left(  5\omega t+5\varphi\right)  $
(cf. Eqs.
S4 and S5, thin solid line ) versus time for four values of the shape
parameter: \textbf{a}, $m=0.5$; \textbf{b}, $m=0.717\simeq
m_{\max}^{impulse}%
$; \textbf{c}, $m=0.9$; \textbf{d}, $m=0.95$.
}
\label{figS2}
\end{figure}

Third, regarding numerical simulations, we considered the entire Fourier
expansion of the elliptic SE in order to obtain useful information concerning
the effectiveness of the approximations used in the theoretical analysis and
experiments (cf. Eqs. S6 and S7, respectively).

\subsection{Chaotic threshold from Melnikov analysis}

Melnikov introduced a function (now known as the Melnikov function (MF),
$M\left(  t_{0}\right)  $) which measures the distance between the perturbed
stable and unstable manifolds in the Poincar\'{e} section at $t_{0}$. If the
MF presents a simple zero, the manifolds intersect transversally and chaotic
instabilities result. See Refs.\cite{S_2,S_3} for more details about MA. Regarding
Eq.~(2) in the main text, note that keeping with the assumption of the MA, it
is assumed that one can write $\delta=\varepsilon\overline{\delta},\gamma=
\varepsilon\overline{\gamma},\eta=\varepsilon\overset{\_}{\eta}$ where
$0<\varepsilon\ll1$ while $\overset{\_}{\delta},\overline{\gamma},\overset
{\_}{\eta},\beta,\omega$ are of order one. Thus, the application of MA to
Eq.~(2) in the main text yields the MF%

\begin{align}
M^{\pm}\left(  t_{0}\right)   &  =-D\pm A\sin\left(  \omega t_{0}\right)
+\frac{\pi\eta}{6\beta}\sum_{p=0}^{\infty}a_{p}\left(  m\right)  b_{p}\left(
T\right)  \cos\left[  \Omega_{p}\left(  T\right)  t_{0}+\left(  2p+1\right)
\varphi\right]  ,\tag{S8}\\
D &  \equiv\frac{4\delta}{3\beta},\tag{S9}\\
A &  \equiv\sqrt{\frac{2}{\beta}}\pi\gamma\omega\operatorname{sech}\left(
\pi\omega/2\right)  ,\tag{S10}\\
\Omega_{p}\left(  T\right)   &  \equiv\left(  2p+1\right)  \frac{2\pi}%
{T},\tag{S11}\\
b_{p}\left(  T\right)   &  \equiv\Omega_{p}^{2}\left(  4+\Omega_{p}%
^{2}\right)  \operatorname{csch}\left(  \frac{\pi\Omega_{p}}{2}\right)
,\tag{S12}%
\end{align}
where the coefficients $a_{p}\left(  m\right)  $ are given by Eq. S5, and
where the positive (negative) sign refers to the right (left) homoclinic orbit
of the underlying conservative Duffing oscillator $\left(  \delta=\eta
=\gamma=0\right)  $:
\begin{align}
x_{0,\pm}\left(  t\right)   &  =\pm\sqrt{\frac{2}{\beta}}\operatorname{sech}%
\left(  t\right)  ,\tag{S13}\\
\overset{.}{x}_{0,\pm}\left(  t\right)   &  =\mp\sqrt{\frac{2}{\beta}%
}\operatorname{sech}\left(  t\right)  \tanh\left(  t\right)  .\tag{S14}%
\end{align}
Let us assume that, in the absence of any SE $\left(  \eta=0\right)  $, the
damped driven two-well Duffing oscillator (Eq. 2 in the main text) presents
chaotic behaviour for which the respective MF,%
\begin{equation}
M_{0}^{\pm}\left(  t_{0}\right)  \equiv-D\pm A\sin\left(  \omega t_{0}\right)
,\tag{S15}%
\end{equation}
has simple zeros, i.e., $D\leqslant A$ or%
\begin{equation}
\gamma\geqslant\gamma_{th}\equiv\frac{2\sqrt{2}\delta\cosh\left(  \pi
\omega/2\right)  }{3\pi\sqrt{\beta}\omega},\tag{S16}%
\end{equation}
where the equal sign corresponds to the case of tangency between the stable
and unstable manifolds \cite{S_3}. If we now let the SE act on the Duffing oscillator
such that $B^{\ast}\leqslant A-D$, with
\begin{equation}
B^{\ast}\equiv\max_{t_{0}}\left\{  \frac{\pi\eta}{6\beta}\sum_{p=0}^{\infty
}a_{p}\left(  m\right)  b_{p}\left(  T\right)  \cos\left[  \Omega_{p}%
t_{0}+\left(  2p+1\right)  \varphi\right]  \right\}  ,\tag{S17}%
\end{equation}
then this relationship represents a sufficient condition for $M^{\pm}\left(
t_{0}\right)  $ to change sign at some $t_{0}$. Thus, a necessary condition
for $M^{\pm}\left(  t_{0}\right)  $ to always have the same sign is%
\begin{equation}
B^{\ast}>A-D\equiv B_{\min}.\tag{S18}%
\end{equation}
Since $a_{p}\left(  m\right)  >0,b_{p}\left(  T\right)  >0,p=0,1,2,...$, one
straightforwardly obtains%
\begin{equation}
B^{\ast}\leqslant\frac{\pi\eta}{6\beta}\sum_{p=0}^{\infty}a_{p}\left(
m\right)  b_{p}\left(  T\right)  ,\tag{S19}%
\end{equation}
and hence,
\begin{align}
\eta &  >\eta_{\min}\equiv\left(  1-\frac{D}{A}\right)  R,\tag{S20}\\
R &  \equiv\frac{6\beta A}{\pi\sum_{p=0}^{\infty}a_{p}\left(  m\right)
b_{p}\left(  T\right)  }.\tag{S21}%
\end{align}
Note that Eq. S20 provides a lower threshold for the amplitude of the SE.
Similarly, an upper threshold is obtained by imposing the condition that the
SE may not enhance the initial chaotic state (i.e., it does not increase the
(initial) gap from the homoclinic tangency condition),%
\begin{equation}
B^{\ast}\leqslant\frac{\pi\eta}{6\beta}\sum_{p=0}^{\infty}a_{p}\left(
m\right)  b_{p}\left(  T\right)  <A+D\equiv B_{\max},\tag{S22}%
\end{equation}
and hence,%
\begin{equation}
\eta<\eta_{\max}\equiv\left(  1+\frac{D}{A}\right)  R,\tag{S23}%
\end{equation}
which is a necessary condition for $M^{\pm}\left(  t_{0}\right)  $ to always
have the same sign. Thus, the suitable (suppressory) amplitudes of the SE must
satisfy%
\begin{equation}
\eta_{\min}<\eta<\eta_{\max},\tag{S24}%
\end{equation}
while the width of the range of suitable amplitudes reads%
\begin{equation}
\Delta\eta\equiv\eta_{\max}-\eta_{\min}=\frac{16\delta}{\pi\sum_{p=0}^{\infty
}a_{p}\left(  m\right)  b_{p}\left(  T\right)  }.\tag{S25}%
\end{equation}
Figures S3 and S4 show how both the width of the range of suitable amplitudes
$\Delta\eta$ (Eq. S25) and the threshold amplitudes $\eta_{\min},\eta_{\max}$
present a single minimum at $m=m_{\min}$ as the shape parameter $m$ is
increased from 0 to 1 due to the dependence of the function $R$ on the shape
parameter. While this minimum $m_{\min}\equiv m_{\min}(T)$ is very near
$m_{\max}^{impulse}\simeq0.717$ over a wide range of periods, one cannot
expect an exact agreement between $m_{\min}$ and $m_{\max}^{impulse}$ for all
periods owing to the dependence of the chaotic threshold on the common
excitation period (main resonance). This means that ever lower amplitudes
$\eta_{\min}$ can suppress chaos as the impulse transmitted by the SE
approaches its maximum value, whereas the corresponding suppressory ranges
$\Delta\eta$ also decrease in the \textit{same }way as $\eta_{\min}$ owing to
the impulse-induced enhancement of the chaos-inducing effectiveness of the SE.
This dependence of $\eta_{\max},\eta_{\min},\Delta\eta$ on the SE's impulse
represents a genuine feature of the impulse-induced chaos-control scenario.
\begin{figure}
\includegraphics[width=0.55\textwidth]{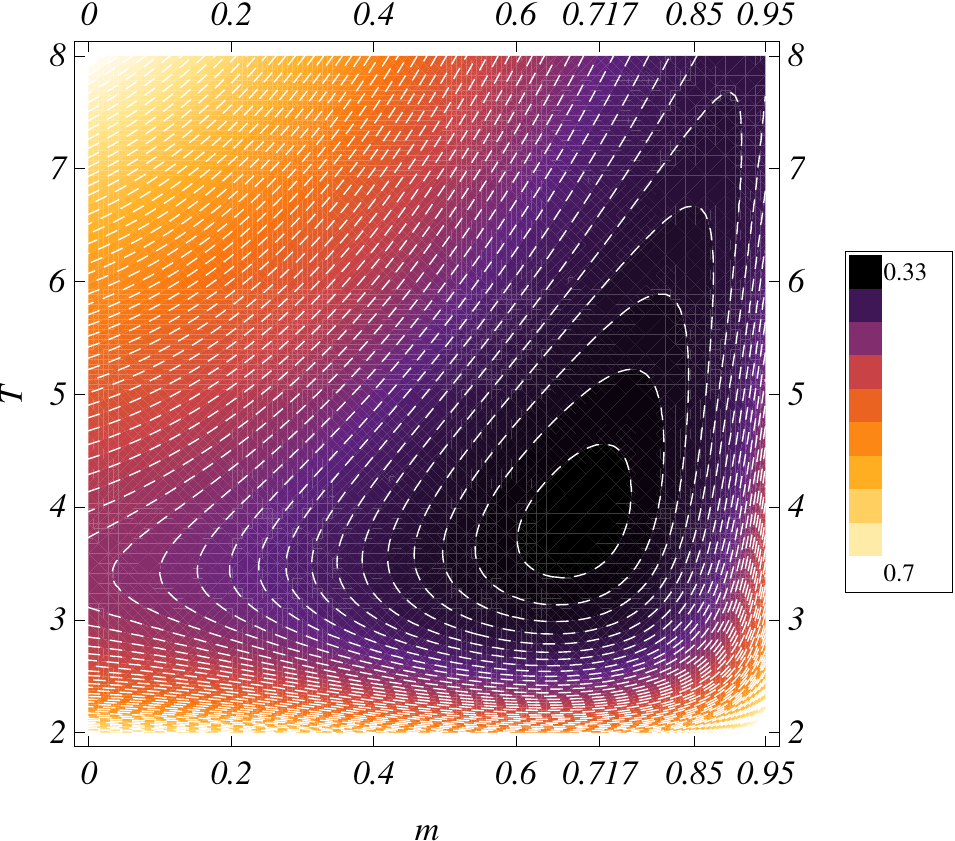}
\caption{\textbf{Width of the range of suitable suppressory
amplitudes in
the }$m-T$ \textbf{control plane.} Contour plot of the function $\Delta
\eta\equiv\eta_{\max}-\eta_{\min}$ (Eq. S24) versus shape parameter
$m$ and
period $T$. Note the existence of an absolute minimum at $m\simeq
m_{\max
}^{impulse},T\simeq4$. System parameters:
$\gamma=0.29,\delta=0.25,\beta=1.$
}
\label{figS3}
\end{figure}
\begin{figure}
\includegraphics[width=0.55\textwidth]{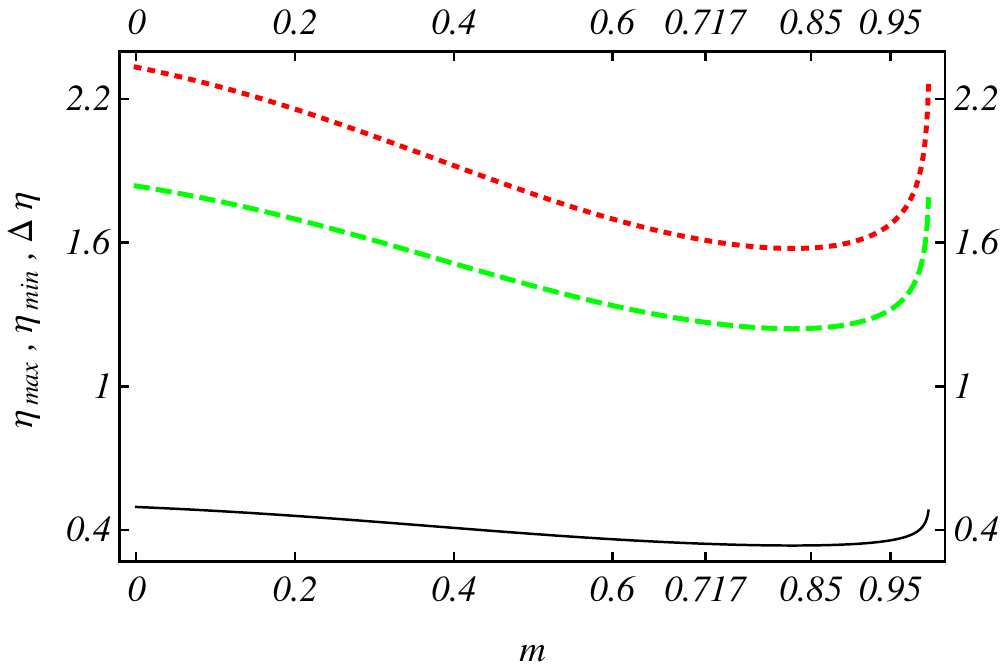}
\caption{\textbf{Threshold amplitudes and width of the
range of suitable
suppressory amplitudes versus shape parameter.} Upper threshold
amplitude
$\eta_{\max}$ (Eq. S22, dotted line), lower threshold amplitude
$\eta_{\min}$
(Eq. S19, solid line), and difference
$\Delta\eta\equiv\eta_{\max}-\eta_{\min
}$ (Eq. S24, dashed line) versus shape parameter $m$. $\omega=1$ and the
remaining parameters as in Fig. S3.
}
\label{figS4}
\end{figure}

Regarding suitable values of the initial phase difference $\varphi$, note that
$\varphi$ determines the relative phase between $M_{0}^{\pm}\left(
t_{0}\right)  $ and
\[
\frac{\pi\eta}{6\beta}\sum_{p=0}^{\infty}a_{p}\left(  m\right)  b_{p}\left(
T\right)  \cos\left[  \Omega_{p}t_{0}+\left(  2p+1\right)  \varphi\right]
\]
irrespective of the shape parameter value. We, therefore, conclude from
previous theory \cite{S_4} that a sufficient condition for $\eta_{\min}<\eta
<\eta_{\max}$ to also be a sufficient condition for suppressing chaos is that
$M_{0}^{\pm}\left(  t_{0}\right)  $ and
\[
\frac{\pi\eta_{\min,\max}}{6\beta}\sum_{p=0}^{\infty}a_{p}\left(  m\right)
b_{p}\left(  T\right)  \cos\left[  \Omega_{p}t_{0}+\left(  2p+1\right)
\varphi\right]
\]
are in opposition. This yields the optimum suppressory values%
\begin{equation}
\varphi_{opt}\equiv\left\{  \frac{\pi}{2},\frac{3\pi}{2}\right\}  \tag{S26}%
\end{equation}
for \textit{all} $m\in\left[  0,1\right]  $ in the sense that they allow the
widest amplitude ranges for the elliptic SE.

To obtain an useful analytical estimate of the boundaries of the regions in
the $\varphi-\eta$ control plane where chaos is suppressed, we assume the
first-harmonic approximation given by Eq. S6 instead of the entire Fourier
expansion (cf. Eq. S4) in the remainder of this section. Indeed, recall that
the value $m_{\max}^{impulse}\simeq0.717$ at which the SE's impulse presents a
single maximum is very close to the value $m=m_{\max}\left(  n=0\right)
\simeq0.642$ where the amplitude $a_{0}\left(  m\right)  $ (cf. Eq. S5)
presents a single maximum (see Fig. S1). Thus, we apply MA to the
\textit{effective} MF%
\begin{align}
M_{eff}^{\pm}\left(  t_{0}\right)   &  =-D\pm A\sin\left(  \omega
t_{0}\right)  +B_{0}\cos\left(  \omega t_{0}+\varphi\right)  ,\tag{S27}\\
B_{0}  &  \equiv\frac{\pi\eta}{6\beta}a_{0}\left(  m\right)  b_{0}\left(
T\right)  , \tag{S28}%
\end{align}
while the effectiveness of the first-harmonic approximation $\left(
\eta>0\right)  $ at suppressing chaos will be examined by considering for
example the effective MF $M_{eff}^{+}\left(  t_{0}\right)  $ (the analysis of
$M_{eff}^{-}\left(  t_{0}\right)  $ is similar and leads to the same
conclusions). To this end, it is convenient to use the normalized MF
$M_{n}^{+}\left(  t_{0}\right)  \equiv M_{eff}^{+}\left(  t_{0}\right)  /D$ to
write%
\begin{align}
M_{n}^{+}\left(  t_{0}\right)   &  =-1+\left(  R^{\prime}-R^{\prime\prime}%
\sin\varphi\right)  \sin\left(  \omega t_{0}\right)  +R^{\prime\prime}%
\cos\varphi\cos\left(  \omega t_{0}\right) \nonumber\\
&  \leqslant-1+\sqrt{\left(  R^{\prime}-R^{\prime\prime}\sin\varphi\right)
^{2}+R^{\prime\prime2}\cos^{2}\varphi}, \tag{S29}%
\end{align}
where $R^{\prime}\equiv A/D,R^{\prime\prime}\equiv B_{0}/D$. If one now lets
the first-harmonic approximation act on the system such that%
\begin{equation}
\left(  R^{\prime}-R^{\prime\prime}\sin\varphi\right)  ^{2}+R^{\prime\prime
2}\cos^{2}\varphi\leqslant1, \tag{S30}%
\end{equation}
this relationship represents a sufficient condition for $M_{n}^{+}\left(
t_{0}\right)  $ to be negative (or null) \ for all $t_{0}$. The equals sign in
Eq.~S30 yields the boundary of the region in the $\varphi-\eta$ plane where
chaos is suppressed:%
\begin{equation}
\eta=\frac{6\sqrt{2\beta}\gamma\tanh\left(  \pi\omega/2\right)  }{a_{0}\left(
m\right)  \omega\left(  4+\omega^{2}\right)  }\left[  \sin\varphi\pm
\sqrt{\frac{\gamma_{th}^{2}}{\gamma^{2}}-\cos^{2}\varphi}\right]  , \tag{S31}%
\end{equation}
with $\gamma>\gamma_{th}$ (cf.~Eq.~S16), and where the two signs before the
square root correspond to the two branches of the boundary (see Fig. S5). The
following remarks may now be in order.

First, the boundary function (Eq. S31) represents two loops encircling the
regularization regions in the $\varphi-\eta$ plane which are symmetric with
respect to the optimal suppressory values
\begin{equation}
\varphi_{opt}\equiv\left\{  \frac{\pi}{2},\frac{3\pi}{2}\right\}  , \tag{S32}%
\end{equation}
respectively, i.e., those values of the initial phase difference for which the
range of suitable suppressory values of $\eta$ is maximum. As expected, they
are the same suppressory values than those found in the exact case of
representing the elliptic SE by its entire Fourier expansion (cf. Eq. S26).

Second, the area, $A_{R}$, enclosed by the boundary function (Eq. S31) is
straightforwardly obtained from previous theory \cite{S_4}:%
\begin{equation}
A_{R}=\frac{32\delta\sinh\left(  \pi\omega/2\right)  }{\pi a_{0}(m)\left(
4\omega^{2}+\omega^{4}\right)  }. \tag{S33}%
\end{equation}
Observe that one finds $A_{R}\rightarrow0$ as $\delta\rightarrow0$, which
corresponds to the limiting Hamiltonian case, as expected. More importantly,
the normalized regularization area
\begin{equation}
\frac{A_{R}(m)}{A_{R}(m=0)}=\frac{a_{0}(m=0)}{a_{0}(m)} \tag{S34}%
\end{equation}
presents, as a function of the shape parameter, a single minimum at the $m$
value where $a_{0}(m)$ presents a single maximum (see Fig. S1): $m_{\max
}\left(  n=0\right)  \simeq0.642$, which is very close to $m_{\max}%
^{impulse}\simeq0.717$. This \textit{inverse} dependence of the regularization
area on the SE's impulse represents a genuine feature of the impulse-induced
chaos-control scenario.

Third, the regularization area shrinks as the ratio $\gamma_{th}/\gamma$
diminishes, which means that the impulse-induced chaos-control scenario is
\textit{sensitive} to the strength of the initial chaotic state in the sense
of its proximity to the threshold condition (cf. Eq. S16).
\begin{figure}
\includegraphics[width=0.55\textwidth]{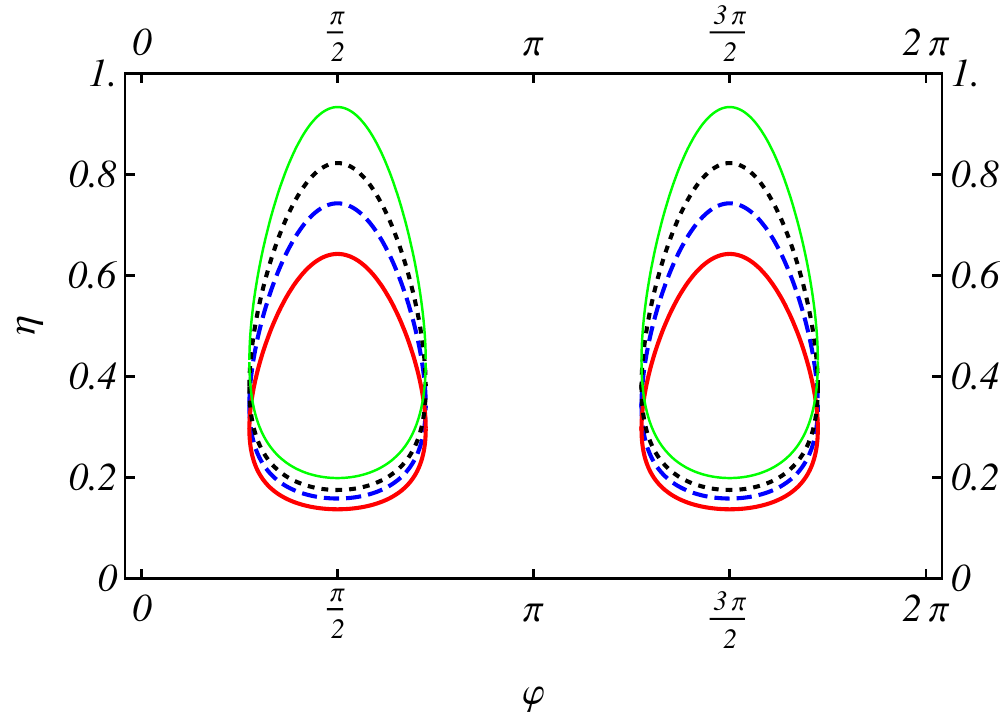}
\caption{\textbf{Analytical estimate of the regularization
boundaries in the
suppressory }$\varphi-\eta$ \textbf{control plane.} Boundary function
(cf. Eq.
S31) encircling the region where chaos is suppressed in the
$\varphi-\eta$
control plane for four values of the shape parameter: $m=0$ (dashed
line),
$m=0.717\simeq m_{\max}^{impulse}$ (solid line), $m=0.93$ (dotted
line), and
$m=0.96$ (thin solid line). System parameters as in Fig. S4.
}
\label{figS5}
\end{figure}
\subsection{Energy-based analysis}

By analyzing the variation of the Duffing oscillator's energy, one
straightforwardly obtains an alternative physical explanation of the foregoing
MA-based predictions. Indeed, Eq.~2 in the main text has the associated energy
equation
\begin{equation}
\frac{dE}{dt}=-\delta\overset{.}{x}^{2}+\gamma\overset{.}{x}\sin\left(  \omega
t\right)  -\beta\eta\overset{.}{x}x^{3}f(t), \tag{S35}%
\end{equation}
where, for the sake of convenience, we introduced the shift $t\rightarrow
t+T/4$, and hence $\varphi\rightarrow\varphi-\pi/2$, and where $E(t)\equiv
\left(  1/2\right)  \overset{.}{x}^{2}\left(  t\right)  +U\left[  x\left(
t\right)  \right]  $ $\left[  U(x)\equiv-x^{2}/2+\beta x^{4}/4\right]  $ is
the energy function. Integration of Eq.~S35 over \textit{any} interval
$\left[  nT,nT+T/2\right]  $, $n=0,1,2,...$, yields
\begin{align}
E\left(  nT+T/2\right)   &  =E(nT)-\delta\int_{nT}^{nT+T/2}\overset{.}{x}%
^{2}\left(  t\right)  dt-\beta\eta\int_{nT}^{nT+T/2}\overset{.}{x}\left(
t\right)  x^{3}\left(  t\right)  f\left(  t\right)  dt\nonumber\\
&  +\gamma\int_{nT}^{nT+T/2}\overset{.}{x}\left(  t\right)  \sin\left(  \omega
t\right)  dt. \tag{S36}%
\end{align}
Now, if we consider fixing the parameters $\left(  \delta,\gamma
,\beta,T\right)  $ for the Duffing oscillator to undergo chaotic behaviour at
$\eta=0$, there always exists an $n=n^{\ast}$ such that the energy increment
$\Delta E\equiv$ $E\left(  n^{\ast}T+T/2\right)  -E(n^{\ast}T)$ is positive
before chaotic escape from one of the two potential wells. Thus, after
applying \ the first mean value theorem for integrals \cite{S_5} together with
well-known properties of the Jacobian elliptic functions \cite{S_1} to the last two
integrals on the right-hand side of Eq.~S36,
\begin{align}
E\left(  n^{\ast}T+T/2\right)   &  =E(n^{\ast}T)-\delta\int_{n^{\ast}%
T}^{n^{\ast}T+T/2}\overset{.}{x}^{2}\left(  t\right)  dt+\frac{\gamma T}{\pi
}\overset{.}{x}\left(  t^{\ast}\right) \nonumber\\
&  -\frac{\beta\eta T\overset{.}{x}\left(  t^{\ast\ast}\right)  x^{3}\left(
t^{\ast\ast}\right)  }{2}F\left(  \varphi,m\right)  , \tag{S37}%
\end{align}
where $t^{\ast},t^{\ast\ast}\in\left[  n^{\ast}T,n^{\ast}T+T/2\right]  $ and
\begin{equation}
F\left(  \varphi,m\right)  \equiv\frac{\sqrt{1-m}N(m)}{K(m)}\operatorname{sd}%
\left[  \frac{2K(m)\varphi}{\pi}\right]  , \tag{S38}%
\end{equation}
with $\operatorname{sd}\left(  \cdot\right)  \equiv\operatorname{sn}\left(
\cdot;m\right)  /\operatorname*{dn}\left(  \cdot;m\right)  $ being the
Jacobian elliptic function of parameter $m$, one has
\begin{equation}
\gamma T\overset{.}{x}\left(  t^{\ast}\right)  /\pi>\delta\int_{n^{\ast}%
T}^{n^{\ast}T+T/2}\overset{.}{x}^{2}\left(  t\right)  dt \tag{S39}%
\end{equation}
at $\eta=0$ when the Duffing oscillator exhibits chaotic behaviour. It is
straightforward to see that $F\left(  \varphi,m\right)  $ presents an absolute
maximum (minimum) at $m=m_{\max}^{impulse}\simeq0.717,\varphi=\pi/2$
($m=m_{\max}^{impulse}\simeq0.717,\varphi=3\pi/2$). It is a $2\pi$-periodic
function in $\varphi$, and presents the noteworthy properties (see Fig. S6):
\begin{align}
F\left(  \pi/2,m\right)   &  =-F\left(  3\pi/2,m\right)  =\frac{N(m)}%
{K(m)}=2I(m),\tag{S40}\\
F\left(  0,m\right)   &  =F\left(  \pi,m\right)  =0,\tag{S41}\\
\lim_{m\rightarrow1}F\left(  \pi/2,m\right)   &  =\lim_{m\rightarrow1}F\left(
3\pi/2,m\right)  =0,\tag{S42}\\
\lim_{m\rightarrow0}F\left(  \pi/2,m\right)   &  =-\lim_{m\rightarrow
0}F\left(  3\pi/2,m\right)  =\frac{2}{\pi}. \tag{S43}%
\end{align}
\begin{figure}
\includegraphics[width=0.55\textwidth]{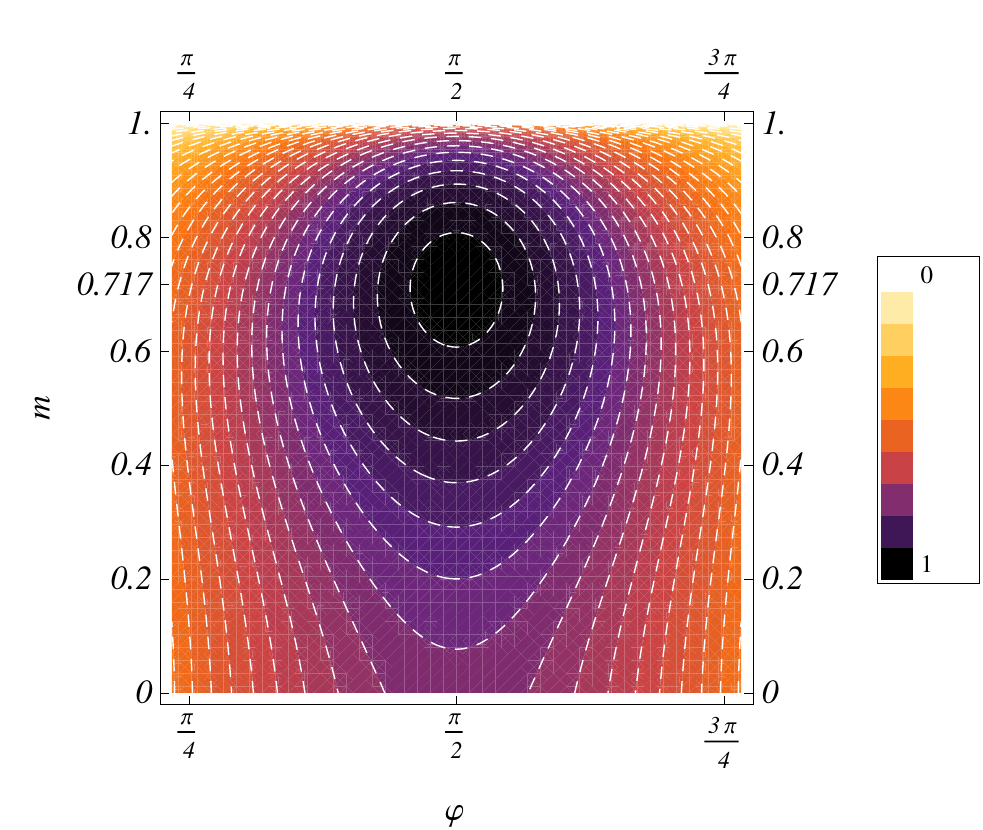}
\caption{\textbf{Function }$F\left(  \varphi,m\right)  $
\textbf{describing
the effect of the SE's impulse in the energy equation.} Contour plot
of the
function $F\left(  \varphi,m\right)  $ (see Eq. S38) versus the
initial phase
difference $\varphi$ and the shape parameter $m$ showing an absolute
maximum
at $\varphi=\varphi_{opt}=\pi/2,m=m_{\max}^{impulse}\simeq0.717$. Note
that
the region around the value $\varphi=\varphi_{opt}=3\pi/2$ is not shown
because of the symmetry $F\left(  \pi/2,m\right)  =-F\left(
3\pi/2,m\right)
=N(m)/K(m)$ (cf. Eq. S40).
}
\label{figS6}
\end{figure}
In this situation, one lets the elliptic SE act on the Duffing oscillator
while holding the remaining parameters constant. For sufficiently small values
of $\eta>0$, one expects that both the dissipation work (the integral in
Eq.~S37) and $\overset{.}{x}\left(  t^{\ast}\right)  $ will approximately
maintain their initial values (at $\eta=0$) while the function $F\left(
\varphi,m\right)  $ will increase (decrease) from 0 (at $\varphi=0,\pi$), so
that, in some cases depending upon the remaining parameters and the sign of
$\overset{.}{x}\left(  t^{\ast\ast}\right)  x^{3}\left(  t^{\ast\ast}\right)
$, the energy increment just before the chaotic escape existing for $\eta=0$,
$\Delta E$, could be sufficiently large and negative to suppress the initial
chaotic state in the sense of leading the Duffing oscillator to the basin of a
certain periodic attractor. Clearly, the probability of suppressing the
initial chaotic state is maximal at $m=m_{\max}^{impulse}\simeq0.717,\varphi
=\pi/2\ \left(  \varphi=3\pi/2\right)  $ (i.e., when the impulse transmitted
by the SE is maximum, cf. Eq. S40), which is in complete agreement with the
foregoing MA-based predictions.

Remarkably, we can obtain an useful alternative estimate of the suppressory
amplitude, $\eta^{\prime}$, by requiring that the sum of the two excitation
terms in Eq. S37 be approximately cancelled:%
\begin{equation}
\frac{\beta\eta^{\prime}\overset{.}{x}\left(  t^{\ast\ast}\right)
x^{3}\left(  t^{\ast\ast}\right)  }{2}F\left(  \varphi,m\right)  \approx
\frac{\gamma}{\pi}\overset{.}{x}\left(  t^{\ast}\right)  . \tag{S44}%
\end{equation}
In such a case, the remaining integral in Eq.~S37 (dissipation work) yields an
energy decrease over time which suppresses the initial chaotic state,
ultimately leading the Duffing oscillator to small-amplitude periodic
oscillations around some of the two fixed points $\left(  x=\pm\beta
^{-1/2},\overset{.}{x}=0\right)  $ of the unperturbed Duffing oscillator
$\left(  \delta=\gamma=\eta=0\right)  $. From the properties of the function
$F\left(  \varphi,m\right)  $ (cf. Eqs. S40-S43), one sees that the lower
values of $\eta^{\prime}$ are obtained for $\varphi=\varphi_{opt}=\left\{
\pi/2,3\pi/2\right\}  $, and hence an alternative estimate of the upper
suppressory amplitude, $\eta_{\max}^{\prime}$, reads
\begin{equation}
\frac{\eta_{\max}^{\prime}(m)}{\eta_{\max}^{\prime}(m=0)}\approx\frac{2/\pi
}{\left\vert F\left(  \pm\pi/2,m\right)  \right\vert }\equiv\frac{2K(m)}{\pi
N(m)}\equiv\left[  \frac{I(m)}{I(m=0)}\right]  ^{-1}, \tag{S45}%
\end{equation}
which presents a single minimum at $m=m_{\max}^{impulse}\simeq0.717$, while
its behaviour, as a function of the shape parameter, is similar to that of the
MA-based upper suppressory amplitude (cf. Eq. S23):
\begin{equation}
\frac{\eta_{\max}(m,T)}{\eta_{\max}(m=0,T)}=\left[  \frac{a_{0}(m)}%
{a_{0}(m=0)}+\sum_{p=1}^{\infty}\frac{a_{p}(m)b_{p}(T)}{a_{0}(m=0)b_{0}%
(T)}\right]  ^{-1}. \tag{S46}%
\end{equation}

\ It is worth noticing that the approximate character of the suppressory
condition given by Eq. S44 prevents us from ensuring that, even in certain
cases corresponding to particular values of the initial conditions and system
parameters, the SE can effectively lead the Duffing oscillator to some of the
two fixed points $\left(  x=\pm\beta^{-1/2},\overset{.}{x}=0\right)  $.
Indeed, Eq. S37 tell us that any decrease of the Duffing oscillator's energy
over half a period implies a subsequent decrease of the dissipation work over
the next half a period, such that this decrease process continues until some
of the mismatches of the (approximate) cancellation of the two excitation
terms is sufficiently large to compensate the dissipation work in the sense of
yielding an increase of the energy, over a certain half a period, and a
subsequent energy oscillation later. This means that the steady behaviour
becomes a small-amplitude periodic oscillation around some of the fixed points
from a certain instant $t=n^{s}T$, while the corresponding dissipation work is
proportional to the action of the periodic orbit in the phase space:%
\begin{equation}
\delta\int_{n^{s}T}^{n^{s}T+T/2}\overset{.}{x}^{2}\left(  t\right)
dt=\delta\int_{n^{s}T}^{n^{s}T+T/2}\overset{.}{x}\left(  t\right)
dx=\delta\pi J, \tag{S47}%
\end{equation}
where $J\equiv\frac{1}{2\pi}\oint pdq$ is the action integral \cite{S_6}.
Alternatively, one can show the same behavior as follows. After linearizing
Eq.~(2) in the main text around $x=\pm\beta^{-1/2}$, one straightforwardly
obtains the equation governing the linear stability of the two equilibria:%
\begin{equation}
\overset{..}{z}+\omega_{0}^{2}z=-\delta\overset{.}{z}-\eta\left(  3z\pm
\beta^{-1/2}\right)  f(t)+\gamma\cos\left(  \omega t\right)  , \tag{S48}%
\end{equation}
where $\omega_{0}\equiv\sqrt{2}$ and $z\equiv x\mp\beta^{-1/2}$, respectively.
Equation S48 has the associated energy equation
\begin{equation}
\frac{dE_{0}}{dt}=-\delta\overset{.}{z}^{2}+\gamma\overset{.}{z}\sin\left(
\omega t\right)  -\eta\left(  3z\pm\beta^{-1/2}\right)  zf(t), \tag{S49}%
\end{equation}
where we introduced the shift $t\rightarrow t+T/4$, and hence $\varphi
\rightarrow\varphi-\pi/2$, and where $E_{0}(t)\equiv\left(  1/2\right)
\overset{.}{z}^{2}\left(  t\right)  +U_{0}\left[  z\left(  t\right)  \right]
$ $\left[  U_{0}(z)\equiv\omega_{0}^{2}z^{2}/2\right]  $ is the energy
function of the linearized system. Integration of Eq.~S49 over \textit{any}
interval $\left[  nT,nT+T/2\right]  $, $n=0,1,2,...$, yields
\begin{align}
E_{0}\left(  nT+T/2\right)   &  =E_{0}(nT)-\delta\int_{nT}^{nT+T/2}\overset
{.}{z}^{2}\left(  t\right)  dt+\gamma\int_{nT}^{nT+T/2}\overset{.}{z}\left(
t\right)  \sin\left(  \omega t\right)  dt\nonumber\\
&  -\eta\int_{nT}^{nT+T/2}\overset{.}{z}\left(  t\right)  \left[
3z(t)\pm\beta^{-1/2}\right]  f\left(  t\right)  dt. \tag{S50}%
\end{align}
After applying \ the first mean value theorem for integrals together with
well-known properties of the Jacobian elliptic functions to the last two
integrals on the right-hand side of Eq.~S50, one obtains
\begin{align}
E_{0}\left(  nT+T/2\right)   &  =E_{0}(nT)-\delta\int_{nT}^{nT+T/2}\overset
{.}{z}^{2}\left(  t\right)  dt+\frac{\gamma T}{\pi}\overset{.}{z}\left(
t^{\prime}\right) \nonumber\\
&  -\frac{\eta T\overset{.}{z}\left(  t^{\prime\prime}\right)  \left[
3z(t^{\prime\prime})\pm\beta^{-1/2}\right]  }{2}F\left(  \varphi,m\right)  ,
\tag{S51}%
\end{align}
where $t^{\prime},t^{\prime\prime}\in\left[  nT,nT+T/2\right]  $. Note that
the suppressory condition given by Eq. S44 implies the approximate
cancellation of the sum of the two excitation terms in Eq. S51, and hence the
same reasoning applied above to the general energy $E$ can now be directly
applied to the small-amplitude energy $E_{0}$ (compare Eqs. S37 and S51), thus
allowing us to conclude that the regularized small-amplitude periodic
oscillations around any of the fixed points $\left(  x=\pm\beta^{-1/2}%
,\overset{.}{x}=0\right)  $ are linearly stable attractors.

\section{Numerical methods}

In our numerical simulations, we studied the purely deterministic case as well
as the robustness of the impulse-induced chaos-control scenario against the
presence of additive noise in the Duffing equation:%

\begin{equation}
\overset{..}{x}=x-\beta\left[  1+\eta f(t)\right]  x^{3}-\delta\overset{.}%
{x}+\gamma\cos\left(  \omega t\right)  +\sqrt{\sigma}\xi\left(  t\right)  ,
\tag{S52}%
\end{equation}
where $\xi\left(  t\right)  $ is a Gaussian white noise with zero mean and
$\left\langle \xi\left(  t\right)  \xi\left(  t+s\right)  \right\rangle
=\delta\left(  s\right)  $, and $\sigma=2k_{b}T^{\ast}$ with $k_{b}$ and
$T^{\ast}$ being the Boltzmann constant and temperature, respectively. For the
sake of completeness, we computed three types of complementary diagrams.

On the one hand, we compare the theoretical predictions obtained from MA with
the Lyapunov exponent (LE) calculations for Eq. S52. In this regard, it is
worth recalling that, even in the case of small values of $\gamma,\delta$ and
$\eta$, one cannot expect too good a quantitative agreement between these two
kinds of approaches because MA is a perturbative technique generally related
to transient chaos, while LE provides information solely concerning steady
responses. We computed the LEs using a version of the algorithm introduced in
\cite{S_7}, with integration typically up to $10^{4}$ drive cycles for each fixed set
of parameters. In the absence of the SE $\left(  \eta=0\right)  $, the Eq. S52
with $\sigma=0,\delta=0.25,\gamma=0.29,\beta=1,\omega=1$ exhibits a strange
chaotic attractor with a maximal LE $\lambda^{+}\left(  \eta=0\right)  =0.025$
bits/s. To construct the LE diagrams we followed two steps. First, the maximal
LE was calculated for each point on a $N\times N$ grid with phase difference
$\varphi$ and amplitude $\eta$ along the horizontal and vertical axes. Second,
a diagram was constructed by only plotting points on the grid according to a
colour code.

On the other hand, we computed period-distribution and isospike
diagrams \cite{S_8}
to obtain detailed information regarding the periodicity order of the
regularized solutions as well as useful information regarding the complexity
of their waveforms in the $\varphi-\eta$ control plane. Isospike diagrams are
based on computing the number of local maxima per period for the periodic
solutions after a sufficiently long transient for each point on a $N\times N$
grid with phase difference $\varphi$ and amplitude $\eta$ along the horizontal
and vertical axes. To this end, after the first $10^{4}$ drive cycles, we
continued the integration for $200$ additional drive cycles recording up to
$800$ extrema (maxima and minima) of the variable of interest and checking
whether pulses repeated or not. In isospike diagrams, black is used to
represent chaos; i.e., lack of numerically detected periodicity. To represent
maxima, we used a palette of 17 colors. Patterns with more than 17 maxima are
plotted by recycling the 17 basic colors modulo 17. Period-distribution
diagrams are based on computing the period of periodic solutions after a
sufficiently long transient ($10^{4}$ drive cycles) for each point on a
$N\times N$ grid with phase difference $\varphi$ and amplitude $\eta$ along
the horizontal and vertical axes. In period-distribution diagrams we used a
colour code to detect periodic solutions with periods between $T$ (period-1
solution) and $8T$ (period-8 solution). In period-distribution diagrams, black
is used to represent chaos; i.e., lack of numerically detected periodicity.
\begin{figure}
\includegraphics[width=0.9\textwidth]{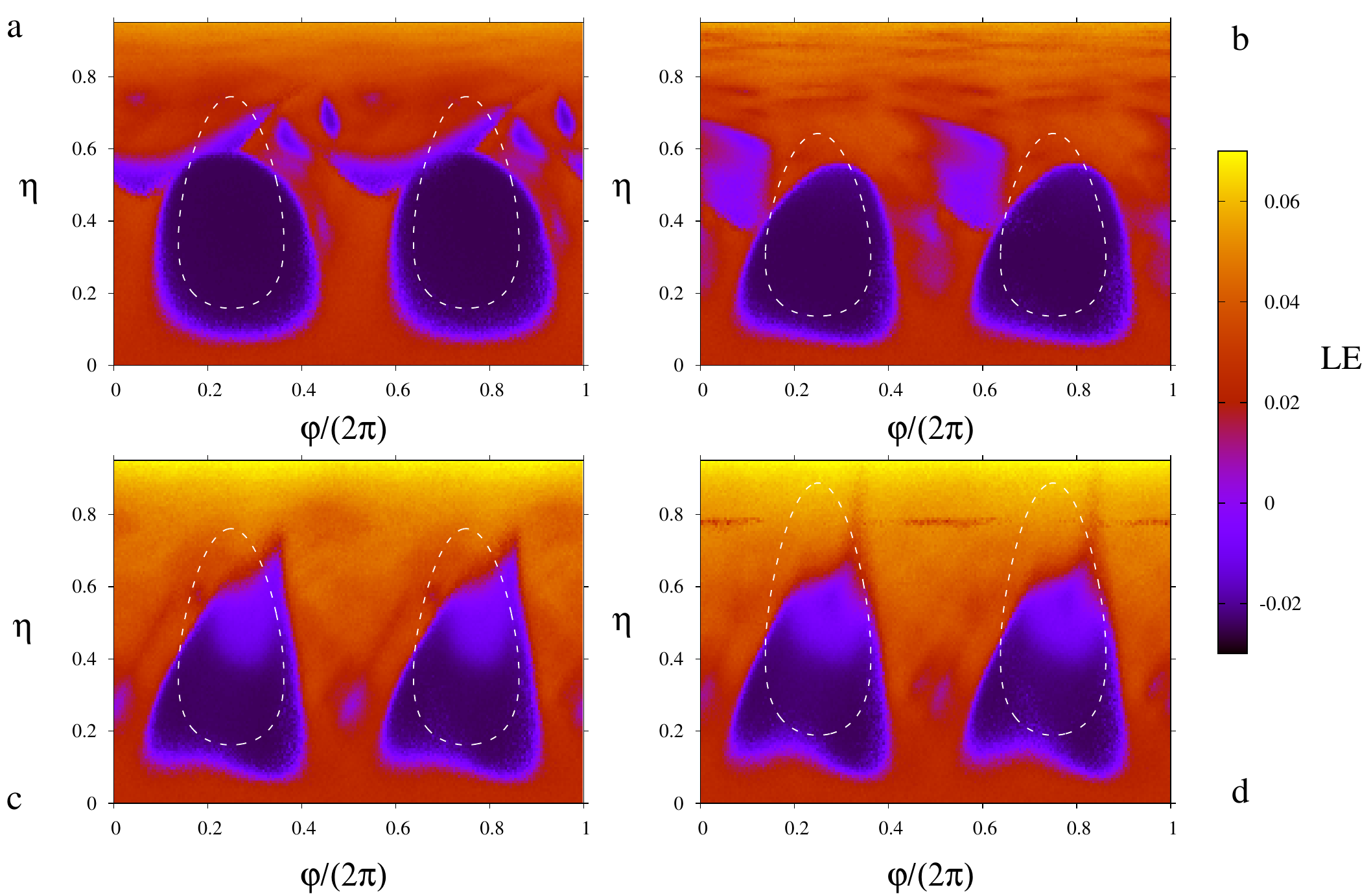}
\caption{\textbf{Robustness of the impulse-induced
chaos-control scenario
against the presence of noise.} LE diagrams in the $\varphi-\eta$
control
plane in the presence of noise for four values of the shape parameter:
\textbf{a}, $m=0$; \textbf{b}, $m=0.717\simeq m_{\max}^{impulse}$;
\textbf{c},
$m=0.9$; \textbf{d}, $m=0.95$. The white contours indicate the
respective
predicted boundary functions for the purely deterministic case
(cf.~Eq. S31)
which are symmetric with respect to the optimal suppressory values of
the
initial phase difference. Noise strength: $\sigma=0.006$, and the
remaining
parameters as in Fig. S4.
}
\label{figS7}
\end{figure}

\begin{figure}
\includegraphics[width=0.9\textwidth]{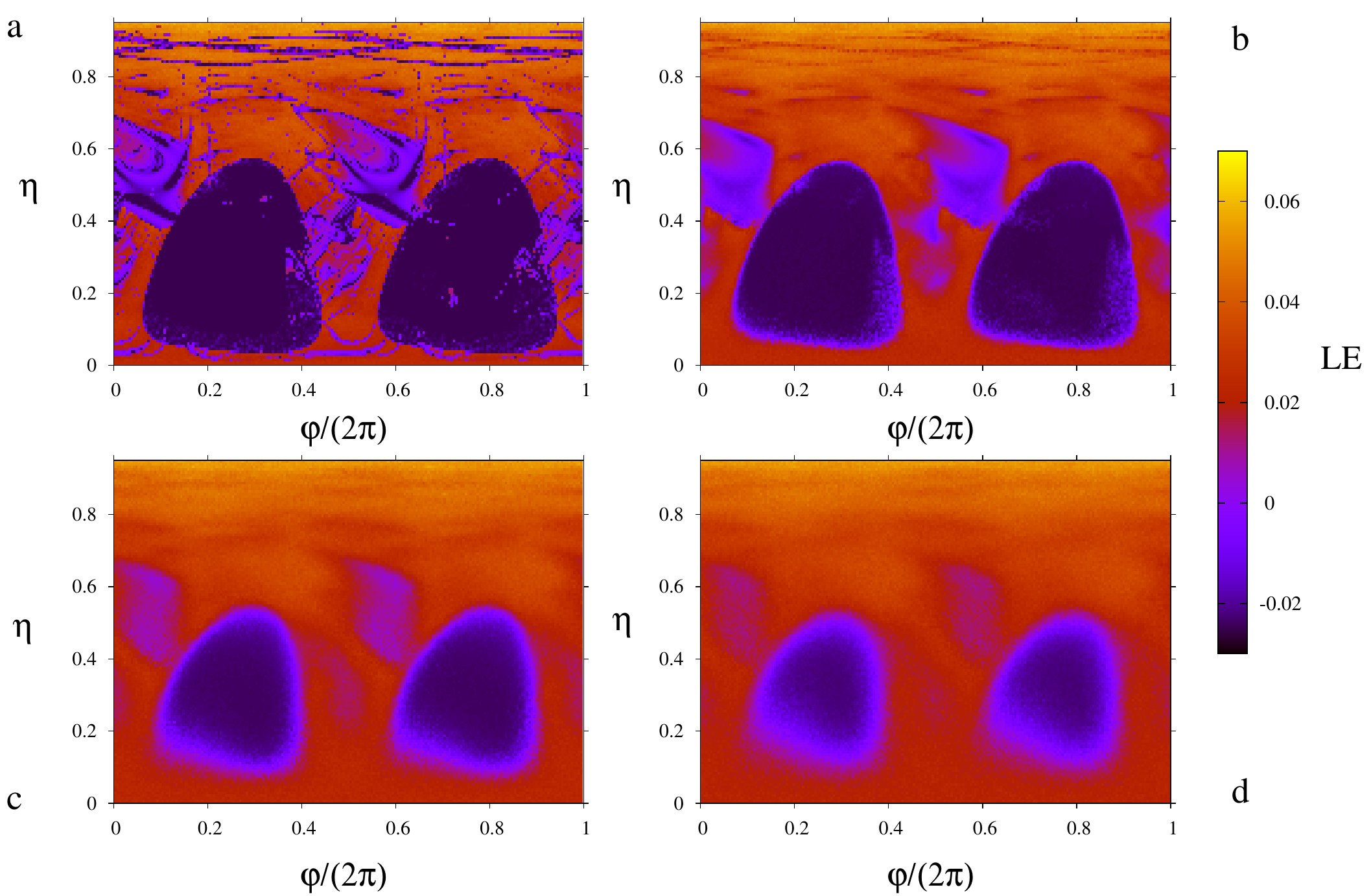}
\caption{\textbf{Robustness of the maximal islands of
regularization against
increasing noise.} LE diagrams in the $\varphi-\eta$ control plane for
four
values of the noise strength: \textbf{a}, $\sigma=0$ (purely
deterministic
case); \textbf{b}, $\sigma=0.002$; \textbf{c}, $\sigma=0.018$;
\textbf{d},
$\sigma=0.038$. Shape parameter: $m=0.717\simeq m_{\max}^{impulse}$,
and the
remaining parameters as in Fig. S4.
}
\label{figS8}
\end{figure}

We studied the evolution of the regularization regions in the $\varphi-\eta$
control plane as the impulse transmitted by the SE is changed from its value
at $m=0$ to its value at an $m$ value very close to $1$ by computing LE,
isospike, and period-distribution diagrams. For the purely deterministic case,
the results are respectively shown in Figs. 4, 5, and 6 of the main text,
while Fig. S7 shows, for the same set of fixed parameters, four illustrative
LE diagrams for the Duffing oscillator in the presence of noise $\left(
\sigma>0\right)  $. Although the presence of noise gives systematically rise
to a decrease, or even a complete elimination, of secondary and minor islands
of regularization in the $\varphi-\eta$ control plane (see Fig. S8), a
comparison between the purely deterministic case $\left(  \sigma=0\right)  $
and the noisy case $\left(  \sigma>0\right)  $ for the same values of the
shape parameter (compare Fig. 4 in the main text with Fig. S7) indicates that
the impulse-induced chaos-control scenario is robust against the presence of
moderate noise.

\section{Experimental methods}

The experimental setup used in our analog implementation of the damped driven
Duffing oscillator (Eq. 2 in the main text) is shown in Fig. S9. The circuit
is governed by the equation%
\begin{align}
\zeta^{-2}\overset{..}{x}  &  =x-\left[  1+\eta f(t)\right]  x^{3}-\zeta
^{-1}\delta\overset{.}{x}+\gamma\cos\left(  2\pi f_{d}t\right)  ,\tag{S53}\\
f(t)  &  \equiv a_{0}(m)\sin\left(  2\pi f_{c}t+\varphi\right)  +a_{1}%
(m)\sin\left(  6\pi f_{c}t+3\varphi\right)  , \tag{S54}%
\end{align}
where $\zeta=\left(  RC\right)  ^{-1}$ with $R=10$ k$\Omega$, $C=10$ nF, while
$\gamma=0.29$ and $f_{d}=1592.500$ Hz are the amplitude and frequency of the
chaos-inducing signal, respectively, $\delta=0.25$, and $f(t)$ is the
two-harmonics approximation of the elliptic SE (cf. Eq. S7). After the
transformation $t\rightarrow\zeta^{-1}t$, Eq. S53 transforms into the
dimensionless Eq. 2 in the main text with $\omega=1$. In the absence of any
elliptic SE $\left(  \eta=0\right)  $, the circuit exhibits steady chaos for
the above set of fixed parameters. The Duffing oscillator block with outputs
$x$ and $y$ which is shown in Fig. S9 has been detailed described in
Ref. \cite{S_9}.
\begin{figure}[h]
\includegraphics[width=0.65\textwidth]{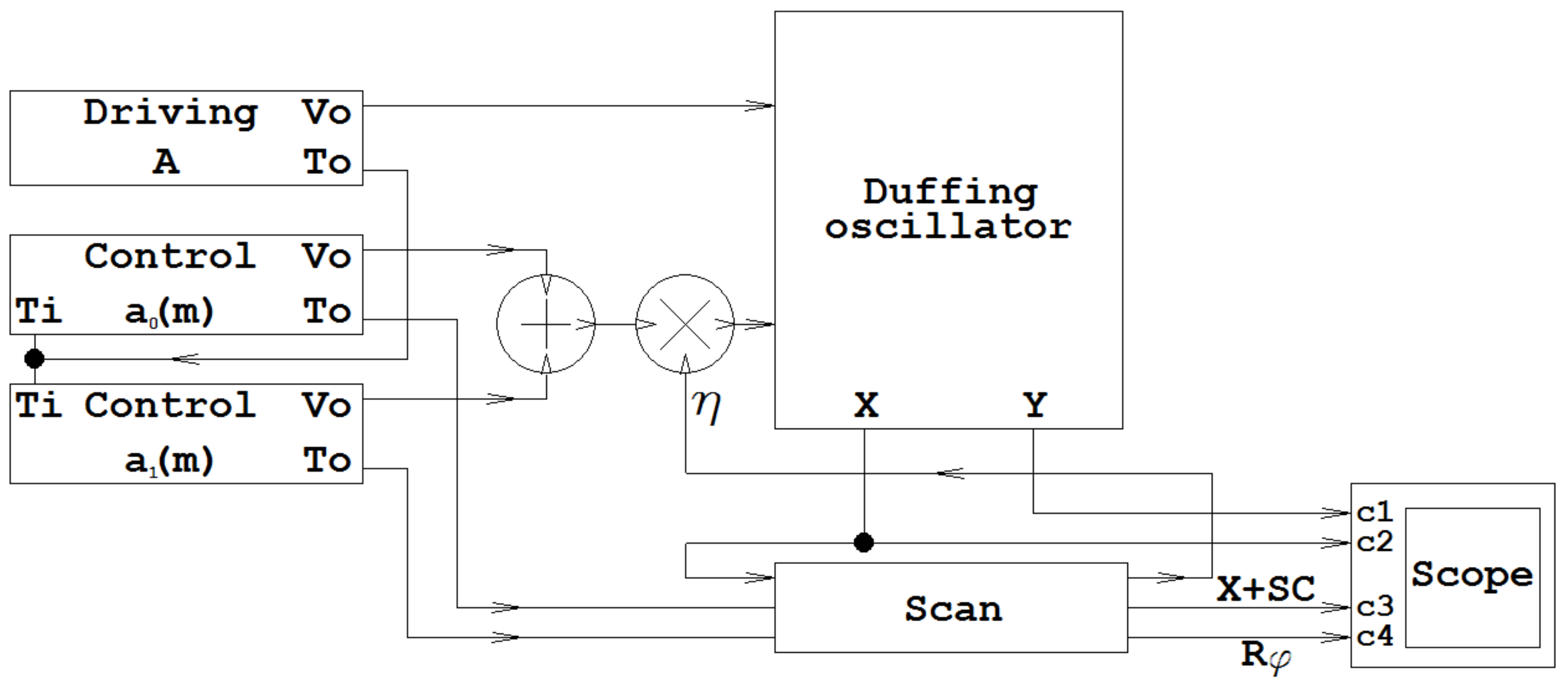}
\caption{\textbf{Scheme of the Duffing's oscillator
circuit.} Blocks diagram
of a damped two-well Duffing oscillator driven by a sinusoidal
chaos-inducing
signal and subjected to an elliptic suppressory signal in the form of a
parametric perturbation of the cubic term. It includes the damped
Duffing
oscillator block with outputs $x$ and $y$, a driving block which
generates the
sinusoidal chaos-inducing signal, while the control blocks generate the
two-harmonics approximation of the elliptic suppressory signal. The
scan block
performs an automatic scanning of the initial phase difference
$\varphi$ and
the suppressory amplitude $\eta$ through the ramp signal $R_{\phi}$
and the
staircase signal $SC$.
}
\label{figS9}
\end{figure}

The initial phase difference $\varphi$ has been implemented by selecting the
frequency of the suppressory signal as $f_{c}=f_{d}+1/T_{sw}$ with $T_{sw}$
being the sweeping phase period during which a phase variation of $2\pi$
occurs, with $T_{sw}=2$ s in the experiments. The scan block generates two
signals: a linear ramp $R_{\phi}$ for a phase variation of $2\pi$ and a $50$
levels staircase signal $SC$ (constant in amplitude during one phase sweep)
allowing us to perform a sweeping of the suppressory amplitude $\eta$. The $x$
and $y$ signals from the Duffing oscillator block together with the phase-ramp
and the $x+SC$ signals are monitored on a four trace oscilloscope.

Unlike the technique used in Ref. \cite{S_10}, where a real-time automatic indicator
was considered to discriminate between regular (periodic) and chaotic
behaviour, we inspect here the temporal series of the $x$ response signal for
each point of the control-plane region $\varphi\in\left[  0,2\pi\right]
,\eta\in\left[  0,1\right]  $ according to the aforementioned resolution. This
procedure provides us not only a reliable discrimination between chaotic and
periodic responses but also to discriminate whether the periodic responses are
low-energy orbits around some of the two fixed points $\left(  x=\pm
\beta^{-1/2},\overset{.}{x}=0\right)  $ of the unperturbed Duffing oscillator
$\left(  \delta=\gamma=\eta=0\right)  $ or higher-energy orbits encircling
both fixed points.

\bigskip
\textbf{Acknowledgements}

P.J.M. and R.C. acknowledge financial support from the Ministerio de
Econom\'{\i}a y Competitividad (MINECO, Spain) through FIS2011-25167 and
FIS2012-34902 projects, respectively. R.C. acknowledges financial support from
the Junta de Extremadura (JEx, Spain) through project GR15146. J. A. C. G. was
supported by CNPq, Brazil.

\end{document}